\documentclass[graybox]{svmult}

\usepackage{mathptmx}       
\usepackage{helvet}         
\usepackage{courier}        
\usepackage{type1cm}        

\usepackage{makeidx}         
\usepackage{graphicx}        
\usepackage{multicol}        
\usepackage[bottom]{footmisc}

\usepackage{booktabs}

\usepackage[caption=false]{subfig}

\usepackage{geometry} 

\makeindex             

\begin{document}

\title*{Epidemic Contact Tracing with Smartphone Sensors}
\author{Khuong An Nguyen, Zhiyuan Luo, Chris Watkins \\ \hfil \\ }
\institute{Khuong An Nguyen \at School of Computing, Engineering \& Mathematics, University of Brighton, UK \\ \email{K.A.Nguyen@bton.ac.uk}
\and Zhiyuan Luo, Chris Watkins \at Computer Science Department, Royal Holloway University of London, UK \\ \email{Zhiyuan.Luo@rhul.ac.uk, C.J.Watkins@rhul.ac.uk}}

%
%
\maketitle

\abstract{Contact tracing is widely considered as an effective procedure in the fight against epidemic diseases. However, one of the challenges for technology based contact tracing is the high number of false positives, questioning its trust-worthiness and efficiency amongst the wider population for mass adoption. To this end, this paper proposes a novel, yet practical smartphone-based contact tracing approach, employing WiFi and acoustic sound for relative distance estimate, in addition to the air pressure and the magnetic field for ambient environment matching. We present a model combining 6 smartphone sensors, prioritising some of them when certain conditions are met. We empirically verified our approach in various realistic environments to demonstrate an achievement of up to 95\% fewer false positives, and 62\% more accurate than Bluetooth-only system. To the best of our knowledge, this paper was one of the first work to propose a combination of smartphone sensors for contact tracing.}

\vspace{1em}






\section{Introduction}
During any viral outbreak, people who have been in close contact with a contagious victim, are at risk of being infected themselves. Therefore, being able to detect such `contacts' early, correctly, and effectively, is critical to manage and suppress the disease. In the past outbreaks (e.g. SARS, Ebola, Swine flu, etc.), contact tracing has proven to be one of the most vital instruments for public health experts. However, as modern viruses (e.g. Covid-19) have evolved to become far deadlier and more infectious, conventional contact tracing approaches are urgently in need to be revamped by modern technology.

In the past decade, the continual proliferation of smartphones has changed the consumers' behaviour. Globally, more than 3.5 billion people own a smartphone, and in the UK alone, more than 94\% of adults have one\footnote{https://www.statista.com/statistics/300378/mobile-phone-usage-in-the-uk - last accessed in 5/2020.}. The smartphone may now be considered as an indispensable necessity to serve most people's daily routines, from essential communications (e.g. family and friend chatting, e-mail and text messaging), to information seeking (e.g. news reading, web surfing, route navigation), to entertainment purposes (e.g. music listening, photo taking, game playing), to health management (e.g. fitness tracking). Coupling with the fact that smartphones are powerful mini-computers equipped with a variety of sensors, this may well be the leverage for technology based contact tracing that we have been searching for.

\begin{figure}[h]
    \centering
    \sidecaption
    \includegraphics[scale=.18]{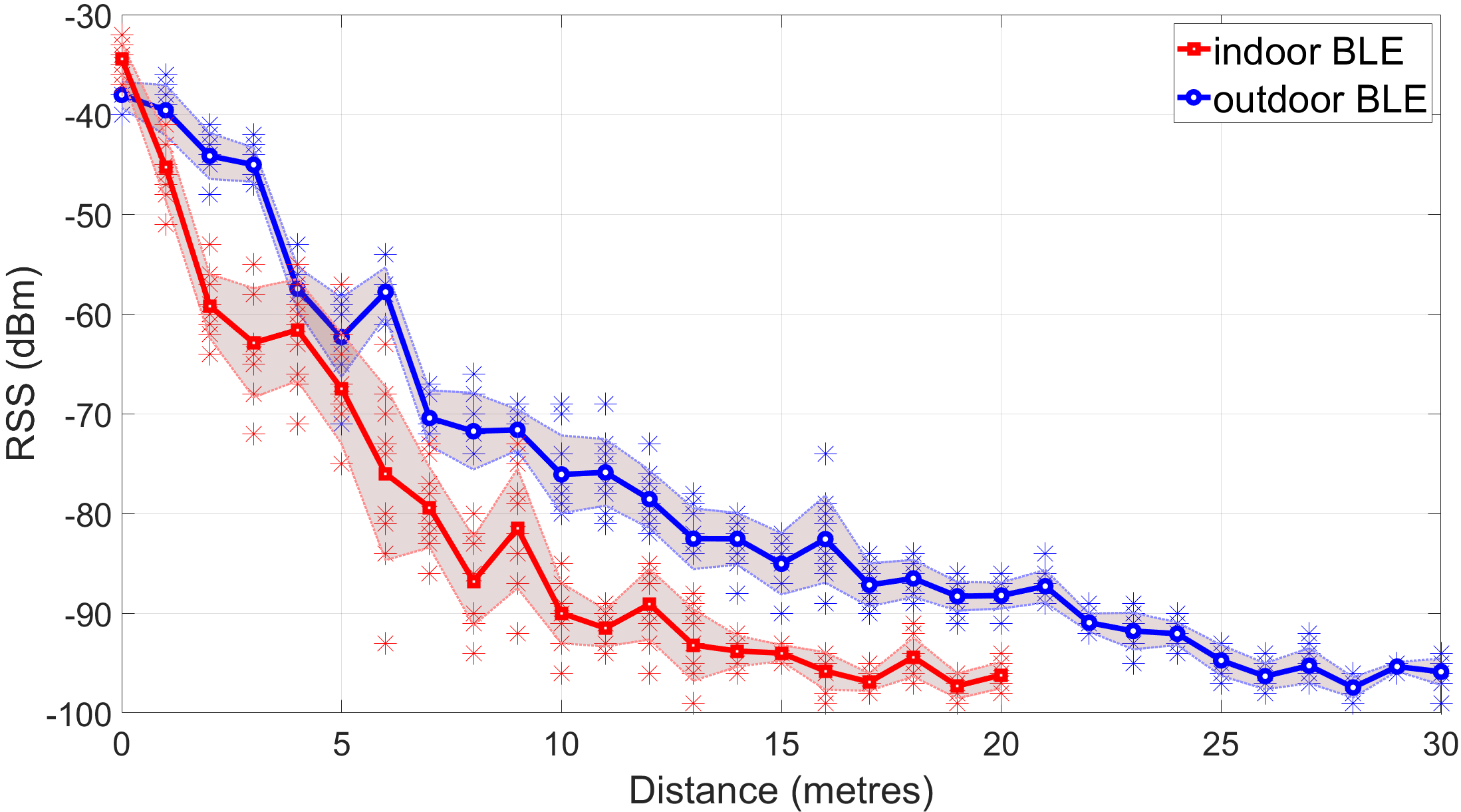}
    \caption{The two problems of BLE based contact tracing system. Firstly, two smartphones may still be visible at 30 metres apart outdoors. Secondly, it is hard to tell the difference between 2 to 6 metres as the RSS standard deviation (demonstrated by the shading areas) is noticeably high. This experiment was performed by measuring the BLE RSS between two phones at pre-determined fixed positions with increasing distance away from each other in a straight line.}
    \label{BLEproblem}       
\end{figure}

Recently, Bluetooth Low Energy (BLE) technology has been viewed as the future prospect for automated contact tracing, thanks to its low power consumption and its relatively short communicating distance. However, from the application viewpoint, BLE bears two major issues, which were revealed by recording the raw BLE received signal strength (RSS) between two smartphones at fixed positions, with increasing distance away from each other in a straight line (see Figure~\ref{BLEproblem}). Visibility wise, two smartphones may still be reached at up to 20 metres indoors and 30 metres outdoors, because of the wireless signal multi-path. Distance wise, it is challenging to determine when two smartphones are 2 metres or 6 metres apart, based on the BLE RSS which varies strongly (much more indoors) because of its frequency hopping technique (to be discussed in Section~\ref{bluetooth}).

To this end, this paper proposes a contact tracing system based on non-location non-intrusive smartphones sensors (no GPS or Cellular). Our approach combines 6 such sensors (i.e. barometer, Bluetooth, magnetometer, microphone, proximity, and WiFi), in one uniform model to detect the contacts of nearby phones. We present extensive empirical experiments to assess the feasibility and performance of such system.

\subsection{The contributions of this paper}
Overall, the paper makes the following novel contributions.
\begin{itemize}
    \item We propose a new concept of multi-smartphone-sensor system for contact tracing.
    \item We assess the feasibility of individual sensor with respect to contact tracing.
    \item We analyse the performance of the proposed system in three real-world testbeds.
\end{itemize}

To the best of our knowledge, this paper was one of the first work to propose a combination of smartphone sensors for contact tracing, and the first one to assess the feasibility and the performance of such approach.

The remaining of the paper is organised into six sections. Section 2 overviews other related work. Section 3 explains the concept of contact tracing, and its significance. So that, Section 4 will build on, to introduce the idea of contact tracing with multiple smartphone sensors. Then, Section 5 details the empirical experiments. Finally, Section 6 concludes our work and outlines future research.

\section{Related work}
As our work is dedicated to co-location smartphone-based contact tracing system, which does not involve any type of location database, we will mainly concentrate on similar approaches in the literature. At the end of the section, Table~\ref{literature} will briefly summarise these existing works.

BLE based contact tracing is perhaps the most popular approach, which was first proposed by the FluPhone project~\cite{yoneki2011fluphone} (note that they also mixed in GPS coordination data). Due to the Covid-19 pandemic, several independent BLE solutions were implemented around the world, often contracted by national agencies (see the MIT list of Covid tracker apps by nations\footnote{https://public.flourish.studio/visualisation/2241702/ - last accessed in 7/2020.}). One of the very first was BlueTrace by the Singaporean government, which senses nearby smartphones via BLE scanning and approximates the distance between them via the BLE RSS~\cite{bay2020bluetrace}, which was also the standard blueprint for most BLE based apps. Most recently, Google and Apple joined force to create the Google/Apple Exposure Notification (GAEN) API\footnote{https://www.apple.com/covid19/contacttracing - last accessed in 7/2020.}, aiming to streamline the BLE background scanning and recording process, as well as addressing the user privacy concern for contact tracing apps, set by the Decentralised Privacy-Preserving Proximity Tracing (DP-3T) protocol~\cite{hubaux2020decentralized}. Unfortunately, it is currently only available for government agencies to adopt, and there have been some concerns of its effectiveness due to sporadic scanning interval and over-simplified BLE scanning report~\cite{dehaye2020swisscovid,leith2020gaen}.

Camera based solution, which is arguably intrusive due to its mechanism of recording of the user's physical appearance (e.g. facial information), has also been proposed in the literature, although very few made it to real-world contact tracing. Some notable examples included the government approaches in India, South Korea, and China~\cite{preethika2020artificial,seetharaman74867177countries,tabari2020nations,vaughan2020tracking}.

GPS based solution, which records the user's precise position in latitude and longitude, has also been attempted, with the most recent work by Raskar et al.~\cite{raskar2020apps}, Wang et al.~\cite{wang2020new}, and the HaMagen app by the Israeli government\footnote{https://govextra.gov.il/ministry-of-health/hamagen-app - last accessed in 7/2020.}~\cite{sonmezdigital}, which build the GPS location trails of the registered users. Having access to this type of detailed location coordination would make contact tracing much easier and more accurate, should the privacy concern be properly addressed.

Magnetometer based approach was proposed in our previous work to track passengers on the public transports~\cite{nguyen2017co}. The two vital observations of this work were that the electric currents powering the rail lines would alter the on-board magnetic field in such a way that people in different carriages experience various non-deterministic measures; and the fact that passengers must share the same journey between at least two consecutive stations. Similar works utilising the magnetic field to detect colocation of the users were reported in other environments, especially indoors with a high degree of magnetic anomalies due to the building infrastructure~\cite{jeon2017judging,jeong2019smartphone,nguyen2019location}. 

WiFi based solution, which features dominantly in the indoor positioning research, has become more attractive for epidemic tracking, thanks to the increasing number of indoor and outdoor public WiFi Access Points (APs). Our previous work demonstrated that contact detection using pure WiFi RSS could closely match the accuracy of GPS (used as a reference) in the city centre, when at least 10 WiFi APs were around~\cite{nguyen2015feasibility}. Similar earlier works focused on the indoor WiFi APs to identify the proximity of the users and the devices~\cite{carlotto2008proximity,krumm2004nearme}.
\begin{table}[h]
	\caption{Overview of existing smartphone-based contact tracing approaches. Most of them are single technology based.}
	\centering
	\begin{tabular}{c c p{10cm}}
		\toprule
		\textbf{Sensor employed} & \textbf{References} & \textbf{Notes} \\
		\midrule
		BLE & \cite{bay2020bluetrace,dehaye2020swisscovid,leith2020gaen,yoneki2011fluphone} & This is the most popular approach for contact tracing right now. It employs BLE scanning for visibility discovery, and the signal strength to estimate the relative distance between devices. \\ \addlinespace[0.2cm]
		
		Camera & \cite{preethika2020artificial,seetharaman74867177countries,tabari2020nations,vaughan2020tracking} & Employing machine learning techniques such as facial recognition to track civilians. \\ \addlinespace[0.2cm]
		
		GPS & \cite{raskar2020apps,sonmezdigital,wang2020new} & Cross-checking the location trails of registered users to discover contacts. \\ \addlinespace[0.2cm]
		
		Magnetism & \cite{jeon2017judging,jeong2019smartphone,nguyen2019location,nguyen2017co} & Relying on the magnetic anomalies caused by ferrous metals in most building infrastructure to record the users' magnetic fingerprints. \\ \addlinespace[0.2cm]
		
		WiFi & \cite{carlotto2008proximity,krumm2004nearme,nguyen2015feasibility} & Utilising the public WiFi APs to detect contacts, where two co-located smartphones share the same set of APs at a particular moment. \\ \addlinespace[0.2cm]
		
		\bottomrule
	\end{tabular}
	\label{literature}
\end{table}

\section{What is contact tracing, and why is it essential ?}
This section introduces the core idea behind contact tracing and its roles in fighting epidemic diseases.

\subsection{What is contact tracing ?}
In essence, contact tracing in an epidemic is the process of identifying all potential victims, who have been in \textit{`close'} contact with an infected person, and iteratively tracing the victims' subsequent contacts in turn.

However, being in close proximity with a contagious individual does not strictly guarantee in getting the virus, as there are other factors such as the person's health condition, the protective equipment (e.g. facial mask) being used, the duration of exposure, the viral load, and many more. The overall hypothesis, adopted by the World Health Organisation (WHO), is that a `contact' is registered when two persons, one of whom is positively tested for Covid-19, are within 1 metre of each other, for at least 15 minutes\footnote{https://www.who.int/docs/default-source/coronaviruse/situation-reports/20200326-sitrep-66-covid-19.pdf - last accessed in 5/2020}.

\subsection{Why we need contact tracing ?}
Most viral infection diseases (e.g. Covid-19, SARS, Swine-flu etc.) share the common trait of being contagious during the incubation period (i.e. the time elapsed between being exposed to the virus, and when the first symptoms are shown) which may last for weeks without any apparent signs, during which the infectious patients unknowingly spread the virus to other victims~\cite{jiang2020does,leung2020difference}. Therefore, it is essential to pro-actively quarantine all potential patients who have been in prolonged contact with the confirmed virus host.

Contact tracing plays three critical roles in the fight against an epidemic. The first role is early treatment, that is, helping exposed patients to seek timely medical treatments, hence boosting successful recovery chance. The second role is transmission control, that is, informing potential victims to self-isolate, hence stopping the chain of onward transmission. The third role is epidemiology study, that is, gaining more insights (e.g. infectious origins, region, route, gender, etc.) of the epidemic, enabling a better strategy to fight the disease in the long term.

\subsection{Three forms of contact tracing}
Conventionally, there are three mainstream forms of contact tracing.
\begin{itemize}
    \item \textbf{Interviews} are perhaps the oldest, but are still widely used in present days, where sick patients are requested to recall as many past contacts as they can~\cite{swanson2018contact}. Such interviews may be completed in person, via a form in the post, or on the internet. Nevertheless, human memory may prove too imprecise for such critical task, not to mention that such interviews are time-consuming and struggle to reach the wider population.
    
    \item \textbf{Narrowcasting} aims at a small-scale, concentrated part of the population (e.g. a town, a building), where the health authorities broadcast an announcement (e.g. on local radio, newspapers, bulletin boards, etc.) asking past visitors to carry out viral tests, and informing others to avoid such areas~\cite{kaligotla2016impact}. This approach allows the officials to quickly control an epidemic hotspot. However, since narrowcasting was meant for targeting a concentrated demographic region, it struggles to stay effective when the epidemic spreads across different regions at large-scale.
    
    \item \textbf{Real-time detection} addresses the issues of both approaches above, by recording the disease contact between citizens and managing the epidemic's progress at it happens. Such detection has mostly been attempted via ambient technology such as facial recognition (via Camera CCTV~\cite{hou2017human,wang2017tracking}), signal tracking (via the phone's cellular signal~\cite{aziz2016smart}), and location monitoring (via GPS data~\cite{chaix2018mobile,olu2016contact,stanley2020limits}).
    
    The common deterrent for all of these technologies is being intrusive, which may lead to the reluctance to comply and adopt by the citizens.
\end{itemize}

In the next section, we will introduce our approach using smartphone sensors that improves on the above real-time detection approach.

\section{Contact tracing with smartphones sensors}
Having instigated the general concept of contact tracing, we may now introduce our approach in smartphone sensors based contact tracing. In doing so, we will outline the idea, process, assumptions, and challenge facing our approach.

\subsection{The smartphone-based contact tracing model}
\label{sec:model}
\begin{figure}[h]
    \centering
    \sidecaption
    \includegraphics[scale=.55]{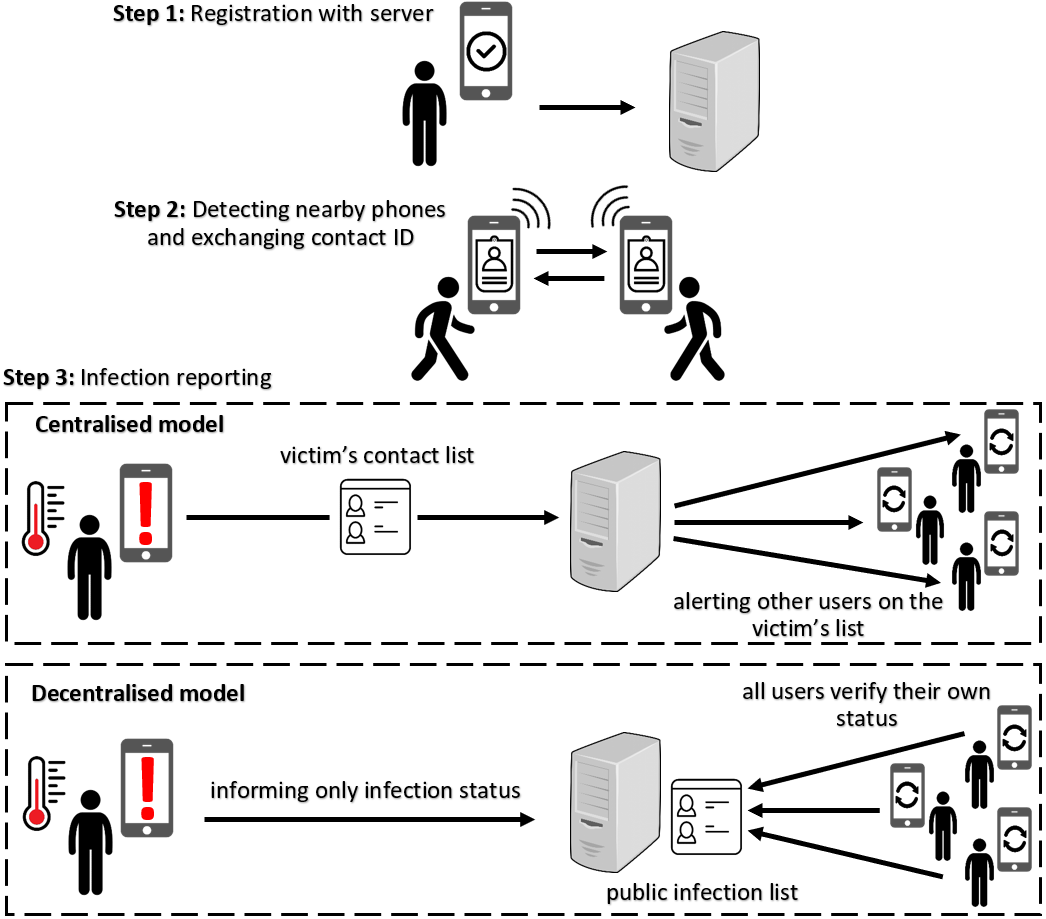}
    \caption{The general steps of our proposed smartphone-based contact tracing procedure.}
    \label{contacttracingprocedure}       
\end{figure}

In principle, our approach fits in the `real-time detection' category, as detailed above. The complete process involves three steps (see Figure~\ref{contacttracingprocedure}).
\begin{itemize}
    \item \textbf{Registration.} All participants download and start the app on their smartphones. It will register with the central server, then generate a unique, temporary ID representing the phone, which will be constantly refreshed after some period of time. The rationale for not using a permanent ID is to make it challenging for snoopers to identify the participants.
    
    \item \textbf{Contact detection.} The app detects nearby phones running the same app, and records such contacts locally on the phone. Both phones will exchange their temporary, current IDs. Although not being the main focus of this paper, the scanning interval should be configured to happen not too frequently (e.g. at 30 seconds or 1 minute interval) to reduce the power consumption and to avoid flooding the BLE and WiFi channels. The rationale for maintaining the contact list locally on each phone, and not centrally on a server is to avoid constant data transmission, preventing potential future data breach, and allowing the non-infectious participants to remain anonymous.
    
    \item \textbf{Infection report.} There are two models for infection report, namely the centralised model, and the decentralised model.
    
    In the centralised model, when a participant is diagnosed positive, his/her app reports the news to a central server, and releasing the locally stored contact list so far. It is clear that such infectious patients must reveal their identity to the server at this stage. The server subsequently informs all contacts from the infected user's list. This model is pioneered by the Pan-European Privacy-Preserving Proximity Tracing (PEPP-PT) group\footnote{https://www.pepp-pt.org - last accessed in 7/2020.}, which promotes standardised approaches (e.g. ROBust and privacy-presERving proximity Tracing protocol - ROBERT~\cite{castelluccia2020robert}) for strong European data privacy in accessing the user smartphone data. Its early adopters included the German and Italian governments~\cite{oxfordeuropean}.
    
    In the decentralised model, the diagnosed participant reports only his/her positive status to the server. S/he still needs to reveal the identity at this point. The server then updates the anonymised public list of infected users, which all participants should frequently check to verify their own status. This model is championed by Google and Apple (i.e. the Google/Apple Exposure Notification (GAEN) API\footnote{https://www.apple.com/covid19/contacttracing - last accessed in 7/2020.}), and the Decentralised Privacy-Preserving Proximity Tracing (DP-3T) initiative~\cite{hubaux2020decentralized}.
    
    The benefit of the centralised setting is that the server's owner (e.g. government or health entities) may have a good overview picture of the epidemic's state, and all infected users may remain anonymous amongst other users (except to the server, and in the rare event of a participant being in contact with just a single user, who is later diagnosed positive). In the opposite, the benefit of the decentralised setting is that an infected user needs not expose his/her entire contact list to the server, giving the participants more control of their own data.
\end{itemize}
 
It is worth noting that we will focus mostly on detecting a contact between two nearby smartphones (i.e. Step 2 in Figure~\ref{contacttracingprocedure}). As such, other important properties such as secured protocol to exchange information, third party trusted server, user privacy, etc. are beyond the scope of this paper.



\subsection{Our assumptions}
In order for smartphones based contact tracing to be effective and reliable, the following assumptions are made:
\begin{itemize}
    \item \textbf{Substantial number of participants.} In common with any other technologies, the success of this approach relies first and foremost to the willingness of the wider population to engage. The first prerequisite is most citizens download the app to their smartphones. This assumption could be satisfied as many countries start to raise their citizen's awareness of the disease's seriousness. It may soon be the pre-condition to relax lock-downs and allow people to better protect themselves.
    
    \item \textbf{Carrying smartphones.} As this approach registers the human contact via the smartphones, it is vital that the devices are present along with the users when such contact happens. This assumption is mostly satisfied as most (if not all) smartphone-owners carry it with them whenever they are outdoors.
    
    \item \textbf{Accurate feedback.} When a user is confirmed positive, s/he must inform the app. Subsequently, ordinary users must not falsify such result. This assumption could be satisfied by official confirmation from the medical test results.
\end{itemize}

It goes without saying that the above assumptions would be made easier, should the proposed technology is shown to be effective with few false positives, and thus gaining trust amongst the users, in which this paper aims to address.

\subsection{Detecting a contact between two smartphones}
Correctly detecting a contact between two nearby smartphones is the key mechanism for this approach. We emphasise that the contact is detected in the proximity based \textit{relative position} (i.e. whether the phones are close or not?), and not the absolute coordination of the phones, which may be too sensitive to privacy issues. The detection happens in real-time, and only a binary (yes/no) decision, along with the nearby phones' temporary ID (as explained in Section~\ref{sec:model}) are recorded locally on the phone.

\begin{figure}[h]
    \centering
    \sidecaption
    \includegraphics[scale=.60]{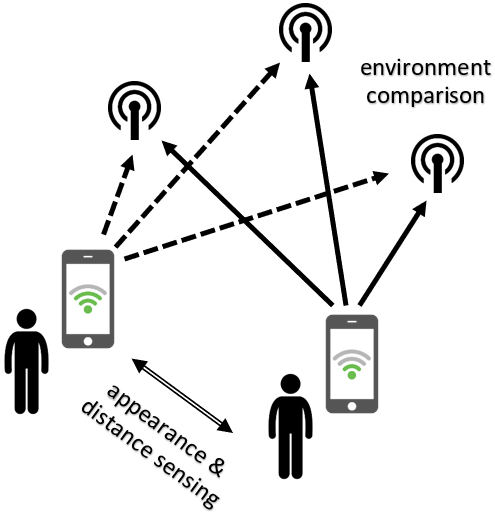}
    \caption{The two approaches for contact tracing with smartphone sensors. Two nearby phones can either sense their appearance and work out the relative distance; or compare the ambient environment they are sharing.}
    \label{approaches}       
\end{figure}

There are two approaches to register such contacts with smartphone sensors (see Figure~\ref{approaches}).
\begin{itemize}
    \item \textbf{Shared environment comparison.} When two phones are nearby, their respective sensor measures of the current environment should be similar.
    
    \item \textbf{Appearance sensing.} Some sensors have the ability to tell the existence of the same sensor type in other nearby phones.
\end{itemize}

\subsection{Assessing the feasibility of employing smartphone sensors}
Currently, there are about 14 sensors in modern smartphones, with different functionalities. Table~\ref{comparisonsensors} compares some of their well-documented properties which are useful for contact tracing, namely the permission, power usage, and sampling rate.
\begin{itemize}
    \item \textbf{Sensor permission.}
    Permission wise, each sensor has different permission levels, based on how Android deems their threat to the user's privacy~\cite{mehrnezhad2019sensor}. In short, there are three relevant types of Android sensor permission for our purpose.
    \begin{itemize}
        \item \textbf{No permission.} No permission is needed to declare anywhere within the app or during run-time. Sensors in this group can silently access the sensors as they wish, which includes the accelerometer, gyroscope, magnetometer, ambient light, and proximity.
        
        \item \textbf{Normal permission.} These sensors pose little risk to the user's privacy. As such, they only need to be declared in the manifest file, and Android will ask the user just once during installation. Once agreed, the user has no way to refuse access in future runs.
        
        \item \textbf{Dangerous permission.} These sensors access high privacy user data (e.g. contact information, location data) or may affect the operation of other apps. For this group of sensors, the app must display a pop-up window explicitly asking the user for permission to access such information, when the app is first launched. Even after accepted, the user has total control to revoke such permission in future runs.
        
    \end{itemize}

    Generally speaking, for the contact tracing purpose, there is a trade-off between usability and privacy. Ideally, we would prefer sensors in the `no permission' or `normal permission' groups as they deliver seamless user experience (e.g. the app works in a simple click without convoluted pop-up messages). Nevertheless, the user should be made clear that such sensors (e.g. Bluetooth, WiFi) have the potential to infer their locations.

    It is worth noting that Android do not simply associate each sensor to a permission type, that is, one sensor may need several permissions, and one permission type may be shared amongst different sensors. For example, when an app needs Bluetooth access, it must specify \textit{BLUETOOTH} permission (for Bluetooth communications), and \textit{BLUETOOTH\_ADMIN} permission (to modify Bluetooth settings) in the `Normal permission' group; as well as the \textit{ACCESS\_FINE\_LOCATION} permission (to initiate a scan) in the `Dangerous permission' group, because Android consider that nearby Bluetooth devices information including the signal strength may be used to indirectly infer the user location.
    
    \item \textbf{Power consumption.} This metric reports the overall energy consumed by the app. In the context of contact tracing, this is an important factor to consider, since it prolongs the battery life between charging cycles, thus allowing more opportunities to detect potential disease contacts, as well as allowing the user to continue with other routines on the phone. An interesting correlation we spotted is that most no permission and normal permission sensors consume little power (see Table~\ref{comparisonsensors}).
    
    \item \textbf{Sampling rate.} This metric is the sensor's frequency in providing the latest measure. High sampling rate sensors are desirable to discover the smallest changes in the environment. Generally speaking, motion related sensors (i.e. accelerometer, gyroscope) have higher frequency than others to detect the quick changes in phone motions. At the other end of the spectrum are Bluetooth, GPS, and Cellular, which will keep scanning and reporting the results as they become available, until the process is interrupted.
\end{itemize}

\begin{table}[!ht]
	\caption{Summary of the relevant properties of 14 common sensors in most smartphones, sorted in alphabetical order. The data are surveyed from the LG G7 ThinQ phone.}
	\centering
	\begin{tabular}{p{2cm}p{1.3cm}p{1.5cm}p{1.3cm}p{1.6cm}p{6cm}}
		\toprule
		\textbf{Sensor} & \textbf{Measure}	& \textbf{Sampling} & \textbf{Power} & \textbf{Permission} & \textbf{Notes}\\
		& \textbf{unit}	& \textbf{rate (max)} & \textbf{usage} & \textbf{type} & \\
		\midrule
		Accelerometer & $m/s^2$ & 500 Hz & low & none & reporting the changing rate of the phone's velocity. \\ \addlinespace[0.2cm]
		
		Ambient light & $lx$ & 4 Hz & low & none & reporting the magnitude of the surrounding light. \\ \addlinespace[0.2cm]

		Barometer & $hPa$ & 120 Hz & low & none & reporting the surrounding atmosphere's pressure. \\ \addlinespace[0.2cm]	
		
		Bluetooth & $dBm$ & various* & medium & dangerous & exchanging information with nearby Bluetooth-enabled devices. \\ \addlinespace[0.2cm]
		
		Camera & image-form & 60 Hz & medium & dangerous & generating an image of the surroundings. \\ \addlinespace[0.2cm]
		
		Cellular & $dBm$ & various* & medium & dangerous & exchanging information with nearby phone towers. \\ \addlinespace[0.2cm]
		
		Fingerprint & image-form & 0.5 Hz & low & normal & generating an image of the human finger. \\ \addlinespace[0.2cm]
		
		GPS	& $s$, $m$ & various* & high & dangerous & reporting the satellite signals, clock timestamp, and status.  \\ \addlinespace[0.2cm]
		
		Gyroscope & $rad/s$ & 500 Hz & low & none & reporting the changing rate of the phone's rotational motion. \\ \addlinespace[0.2cm]
		
		Magnetometer & $\mu T$ & 200 Hz & low & none & reporting the magnitude of the surrounding magnetic field. \\ \addlinespace[0.2cm]
		
		Microphone & $dB$ & 48 kHz & medium & dangerous & reporting the magnitude and the raw surrounding acoustic noise. \\ \addlinespace[0.2cm]
		
		NFC & N/A & 1 Hz & low & normal & exchanging information with nearby RFID tags within 10 cm. \\ \addlinespace[0.2cm]
		
		Proximity & $cm$ & 4 Hz & low & none & reporting the distance to the nearest object within 10 cm. Some proximity sensors only report a binary near/far result. \\ \addlinespace[0.2cm]
		
		
		
		
		WiFi & $dBm$ & 0.03 Hz & high & dangerous & exchanging information with nearby WiFi-enabled devices. \\ \addlinespace[0.2cm]
		\bottomrule
	\end{tabular}
	
	$*$ The phone will continue scanning for nearby Bluetooth devices, GPS satellites and Cellular towers, updating the results as they come in, until the process is interrupted.
	\label{comparisonsensors}
\end{table}

When it comes to picking which sensors to use, we impose a strong criterion that any selected sensor must not directly reveal the smartphone's position, for our sole purpose of proximity tracking, and to avoid potential information misuse. This rules out GPS (which directly yields the longitude and latitude), and Cellular (whose cell tower location database is publicly available\footnote{https://opencellid.org - last accessed in 5/2020.}). The ambient light sensor which measures the lighting intensity, is an interesting source of information. Yet, its readings are too volatile under different phone's angles. On the same note, the camera, fingerprint, NFC, and time-of-flight's usage were rather specific, and do not appear to be useful for our contact tracing purpose at this stage.

Last but not least, as mentioned in the previous section, only the nearby phones' IDs are stored locally on the smartphones, and not the detailed sensor data.


\subsubsection{Bluetooth for proximity detection}
\label{bluetooth}
On the smartphones, Bluetooth technology, with its latest iteration called Bluetooth Low Energy (BLE), was intended to connect the phone to small peripherals (e.g. headphone, gamepad, etc.) in short distance (typically no more than 10 metres, and ideally within 2 to 3 metres), which is a great fit for close contact detection.

The detection process using BLE works as follows. First and foremost, one phone must act in the `Central' role, and the other phone plays the `Peripheral' role. The peripheral phone will constantly sending out unsolicited messages (including its name, MAC address, etc.) on the 40 BLE channels to inform its existence~\cite{contreras2017performance,de2017study,faragher2015location,kalbandhe2016indoor}. The central phone, at any time of preference, will initiate a scan to look for those peripheral phones. Secondly, both phones may establish a BLE connection to exchange information. Ideally, the smartphones will alternate between these two roles to avoid the situation where everyone is listening while no-one is broadcasting, and vice-versa.

It is worth noting that, as a BLE scan will also reveal the received signal strength (RSS), which roughly indicates how far away the nearby peripheral phones are, we could estimate the distance using the well-known inverse square law of Physics, that is, the signal intensity is inversely proportional to the square of the distance from a signal transmitter~\cite{goldsmith2005wireless}.
\begin{equation}
    d = 10^{(\frac{Power - BLE\_RSS} {10 * n})}
\end{equation}
where d is the estimated distance, Power is the RSS measured at 1 metre, $n$ is the signal propagation constant (e.g. $n = 2$ for free-space path loss), BLE\_RSS is the RSS received by the central phone. Although there are other distance models, without knowing how the BLE signal propagates (especially challenging indoors), we opted for this simple form of free-space path loss model.

Nevertheless, there are two challenges with BLE. Firstly, BLE wireless signal does easily penetrate walls and furniture, which contributes to the false positive contact detection (i.e. two neighbours separated by a thick wall may be registered as a contact). Secondly, the correlation between the RSS and the distance is not strictly linear, because of the signal attenuation (i.e. without an unobstructed line-of-sight between the two phones, the signal waves are strengthened or weakened as they travel in different directions in the air), and the BLE frequency hopping technique (i.e. the phone frequently switches between the BLE channels to avoid signal collision, which inadvertently impacts the receiving signal at the other end).

In short, we will only employ BLE as a rough indicator to discover nearby phones (which could be anywhere up to 30 metres in the vicinity), and to kick-start the upcoming procedures involving other sensors. We will later demonstrate empirically in detail how frequency hopping may affect the accuracy of BLE in estimating the relative distance between the phones in Section~\ref{experiments}.

\subsubsection{WiFi for distance measuring}
\label{WiFi}
On the smartphones, WiFi technology was originally intended to connect the phone to the Access Points (APs) for internet access. Recently, the WiFi Direct peer-to-peer protocol enables two WiFi-enabled devices to communicate directly without an AP, over much longer distance (up to 50 metres indoors) than Bluetooth~\cite{khan2017wi}, in which the smartphones will negotiate directly between themselves to automatically assign the central and peripheral role.

From the contact detection's viewpoint, WiFi technology may be employed for environment comparison, appearance sensing, as well as distance measuring purposes.
\begin{itemize}
    \item \textbf{For environment comparison.} The observation is that most modern buildings and public venues have plenty of WiFi APs to provide internet access to residents and customers. As such, two smartphones with a similar set of observed APs are potentially nearby.
    
    \item \textbf{For appearance sensing and distance measuring.} This is performed in a similar fashion as with BLE.
\end{itemize}

However, there are five major challenges for employing WiFi. Firstly, since Android 9, all apps may only initiate a scan at most 4 times every 2 minutes (about once every 30 seconds)\footnote{https://developer.android.com/guide/topics/connectivity/wifi-scan - last accessed in 5/2020}, whilst there is no scanning restriction for Bluetooth\footnote{Strictly speaking, since Android 8, the \textit{startScan()} method does impose four predefined scanning configurations, where the fastest $SCAN\_MODE\_LOW\_LATENCY$ option only scans for about 300 ms. However, it is still possible (even on the latest Android 10) to call the deprecated $startLeScan() / stopLeScan()$ methods introduced since Android 5 to fully control the scanning cycle.}. Secondly, the current implementation of WiFi Direct on Android does not expose the RSS of the peer devices, which means the phone needs to be set up as a WiFi hotspot for its RSS to be harvested via the usual WiFi scan. Thirdly, the glaring weakness of WiFi technology is its long broadcasting distance of up to 50 metres. As such, smartphones on different floors, or even separate buildings may still see each others, or observe the same set of WiFi APs, which invalidates the environment comparison approach. Fourthly, WiFi consumes much higher battery than BLE and other sensors. Lastly, WiFi signals do suffer from the same signal multi-path problem as BLE.

Taking into account these concerns, WiFi should only be employed to complement BLE. In particular, it should only be called into action when a potential contact has been confirmed by BLE. Given the WiFi RSS is considerably more stable (i.e. no frequency hopping as in BLE), its distance conversion may be more accurately represented (to be empirically assessed in Section~\ref{experiments}).

Without loss of generality, given the WiFi RSS sequence (which reflects the distance) between the two smartphones $W=(w_1, \dots, w_N)$, recorded over a time window (e.g. 15 minutes based on the WHO recommended infectious duration\footnote{https://www.who.int/docs/default-source/coronaviruse/situation-reports/20200326-sitrep-66-covid-19.pdf - last accessed in 7/2020}), where $w_i$ $(1 \leq i \leq N)$ is the WiFi RSS at time point $i^{th}$, we employed the free-space path loss model, described in Section~\ref{bluetooth}, to covert the RSS into a distance estimate. The mean value of all estimates within the entire time window will decide if the two phones were within the infection range (e.g. 1 metre based on the WHO guideline).

\subsubsection{Microphone for proximity detection}
\label{sound}

Sound is generated by the vibrations of air particles, which are then picked up by the human ear and the smartphone's microphone~\cite{murakami2018smartphone}. For contact detection, sound may be leveraged for both the appearance sensing and the distance measuring purposes as follows (although it is possible to encode information within those sounds, yet, for our contact tracing purpose, we only need to detect the appearance and the rough distance measure).

For appearance sensing, one smartphone will play the peripheral role by emitting a chirp via its built-in loud speaker. The other phones which act in the central role will pick up those sounds by their built-in microphone, and response with their own chirps to make contact, which indicates that they are in close proximity.

For distance measuring, there are two options. The first one is based on the concept of time difference of arrival. The second one is based on sound amplitude.
\begin{itemize}
    \item \textbf{Time difference of arrival (TDoA).} With this approach, the distance between the two phones is inferred by timing the moment the chirp was sent by one phone, until it is received by the other end. If the clock of both phones are perfectly synchronised, the distance is simply calculated as $distance\;=\;speed\;of\;sound\;*\;elapsed\;time$, with speed of sound is a constant (e.g. 343 $m/s$ at 23 $^{\circ}$C room temperature)~\cite{yavuz2015measuring}.
    
    The rationale of this approach is that acoustic signal travels at a much slower velocity (i.e. 343 $m/s$), compared to WiFi and Bluetooth signals which travel at the speed of light (i.e. 300,000 $km/s$). Therefore, it is more feasible to time and compute such distance.
    
    However, there are several challenges. Firstly, it is unlikely various smartphones would share the same clock timing. Secondly, all Android sound packages (i.e. SoundPool, AudioTrack and OpenSL ES) have considerably non-deterministic latency (i.e. there are unpredictable delay from the moment the audio play command was issued, until the actual sound was sent out by the built-in speaker), in the region of [180 $ms$ to 300 $ms$]. Given that sound travels at 343 $m/s$, an average of 200 $ms$ error in TDoA estimation will lead to more than 70 metres error in ranging estimation, which is simply not usable for our purpose.
    
    \item \textbf{Sound amplitude.} This approach relies on the same concept as in BLE and WiFi RSS ranging, that is, using sound amplitude as distance indicator. The central phone plays a chirp. As this chirp travels in the air, it loses pressure for which the receiving phone may use to work out the distance to the central phone.
\end{itemize}


The final task is designing a chirp signal to be received reliably within short distance by other nearby smartphones, for which there are three criteria to consider.
\begin{itemize}
    \item \textbf{Frequency}. The frequency (in $Hz$) is the speed of vibration which determines the sound pitch (e.g. female voice is perceived to have higher pitch than male one). Distance wise, low frequency sound travels further than high frequency one, as there is less energy being lost in the process. As such, we would prefer high frequency chirp for short distance contact detection. Nevertheless, high frequency chirp tends to attenuate more heavily than low frequency one, because of higher viscosity caused by many peaks pressuring against the air~\cite{hoppe2012acoustic}. The major constraint is that, although our test phones are capable of recording very high frequency acoustic sound, streamlined smartphone microphones may be limited to lower frequencies in the human audible range (i.e. 20 $Hz$ to 20 $kHz$), which was primarily what they were designed for. As such, our chirp signal should be within this range.
    
    \item \textbf{Amplitude.} The amplitude (in $dB$) is the size of the vibration which determines the loudness of the chirp. A high amplitude means louder sound, which in turns, translates to longer travelling distance. For contact detection, we would prefer low amplitude sound for quiet operation as well as limiting the distance to within the short contagion range.
    
    \item \textbf{Duration.} The duration (in $ms$) is the length of the vibration. For contact detection, the chirp duration should be short, as multiple chirps may be sent simultaneously by other nearby phones, and potentially confuse the receiving ends.
\end{itemize}


Taking all of these constraints into consideration, an ideal chirp for contact tracing purpose should have a frequency within 2 $kHz$ to 6 $kHz$ range (as most acoustic sounds start to attenuate greatly above 8 $kHz$~\cite{peng2007beepbeep}), with an amplitude of about 20 $dB$ (still within the recording range of all smartphone's microphones, yet is perceived as just a small whisper for the human ear), and a 50 $ms$ duration (the ideal length to avoid collision with other chirps)~\cite{lazik2012indoor}. Lastly, each smartphone should use a different frequency when communicating to better differentiate with other nearby phones\footnote{Although it is possible to construct a new chirp on the fly with new properties to quickly adjust to the real-time environment, we reckon this process would add unnecessary complexity and power usage to the system. Hence, we leave this for future investigations.}.


There are two advantages of sound based approach. Firstly, while radio waves propagate seamlessly through space, sound waves require a material medium (e.g. water, air) to convey from one place to another, which does not easily penetrate thick walls, furniture. Secondly, we may manipulate the frequency and loudness properties of the a chirp, which indirectly control the travelling distance, whereas although theoretically possible to do the same with BLE, there is no Android API to modify the BLE sensor's sensitivity at the time of writing. Hence, sound is a great fit for contact tracing.

Nevertheless, there are three challenges for employing sound. Firstly, it does suffer from the same multi-path issue as in BLE and WiFi, in which the acoustic signals reach the receiving microphone in different paths, due to reverberation. Secondly, the sound propagation speed does vary according to different temperatures and humidity (this issue may be negligible in real practice, as the distance between the phones is rather short for our purpose). Lastly, while BLE and WiFi encode their unique sender ID within the signal, we need to design such handshaking process from scratch for sound. Our solution is leveraging BLE signal to make contact between the two smartphones first, before transmitting the chirp.

\subsubsection{Magnetometer for ambient magnetic field comparison}
\label{sec:magnetometer}
Magnetism exists everywhere on Earth, which is caused by the movements of molten metal at the Earth's core. This natural magnetic field is strongly perturbed by ferrous metal from most building's materials~\cite{kim2016novel,kim2017indoor,storms2019magnetic}. Therefore, small indoor areas within the building may report different magnetic signatures, which is an opportunity for matching smartphones contact.

The magnetometer measures the ambient magnetic field strength around the phone in 3-dimensional space. However, as the sensor is fixed within the phone body, its coordinates align with the phone's body frame. As such, the orientation of the phone varies the magnetometer reading, even in the same spot. This challenge may be addressed by ignoring the direction of the magnetic field vector, and computing the total scalar magnitude $m$ as follows~\cite{nguyen2017co}.
\begin{equation}
    m = \sqrt{m_x^2 + m_y^2 + m_z^2}
\end{equation}
where $m_x$, $m_y$ and $m_z$ in $(\mu T)$ are the magnetic field strength along the $x$, $y$ and $z$ axis respectively.

Without loss of generality, given Alice's magnetometer measures $A=(m_1, \dots, m_N)$ and Bob's magnetometer measures $B=(m_1', \dots, m_M')$, where $N$ and $M$ may be different due to various sensor's sampling rates. We will apply the method described in~\cite{nguyen2017co} using Dynamic Time Warping (DTW) which was designed to compare the magnetic sequences of different length, and to handle the various sensitivities from different sensor vendors (i.e. a more sensitive sensor may return higher reading values). In essence, DTW finds the optimal warped path between the two sequences by building an N-by-M matrix, in which $[i^{th}, j^{th}]$ is the distance between measures $m_i$ and $m_j'$ calculated as follows~\cite{nguyen2019realtime}.
\begin{equation}
    d(m_i, m_j') = (m_i - m_j')^2
\end{equation}

The optimal warped path of length k : $w(1,1), \dots, w(n,m)$ between the two sequences, that minimises the warping cost is computed as follows~\cite{nguyen2019realtime}.
\begin{equation}
    \hspace{130pt}
	w(n, m) = d(m_i, m'_j) + min\left\{\hspace{5pt}
	\begin{tabular}{@{}l@{}}
	w(i-1, j) \\
	w(i-1, j-1) \\
	w(i, j-1)
	\end{tabular}\hspace{2pt}
	\right\}\hspace{130pt}
\end{equation}

with $(1 < i < N, 1 < j < M)$.

At the end, the DTW score is computed as $w(n,m)$ divided by the length of the warped path. If the computed score of the two magnetic sequences is below the threshold, the two phones are deemed to be in close proximity (see Section~\ref{sec:environment} for the empirical experiments with the magnetic field threshold).

\subsubsection{Barometer for air comparison}
As most viral diseases are airborne, having the ability to detect if two persons are breathing the same air is profoundly valuable. Regrettably, all sensors discussed so far inherit the same weakness from the wireless signal property, that is, electromagnetic and sound wave can penetrate walls. As such, two persons fully segregated by a thick, concrete wall may still be detected by BLE, WiFi, or sound signal, which result in a false positive being registered.

The barometer which measures the ambient air pressure around the phone, may offer a solution for this challenge. The observation is that, the readings are varied by the air weight in the atmosphere, caused by different altitudes (e.g. on different building floors), or by the winds (e.g. a closed indoor space has different measures from an open outdoor one)~\cite{kim2017floor,li2013using,nii2017bar}.

Nevertheless, if the phone is kept in a tight pocket, handbag, etc., the air measures are potentially different, although the two persons are in the same place. This challenge may be addressed with the help of the proximity sensor, which was originally designed to measure the distance to the nearest object facing the phone screen (i.e. it is intended to detect the side of the human face to switch off the screen while answering a call). By using the proximity sensor, the system may detect whether the phone is left in tight space or in the open to trigger the barometer reading.

Without loss of generality, given the barometer measure sequence from two smartphones, we employed the same DTW approach as in the above Section~\ref{sec:magnetometer}, to work out a score reflecting the difference between the two barometer sequences. If the score is below the threshold, the two phones are deemed to be in close proximity, from the barometer's perspective (see Section~\ref{sec:environment} for the empirical experiments with the air pressure threshold).


\subsection{Fusion of sensors information}
\begin{figure}[h]
    \centering
    \sidecaption
    \includegraphics[scale=.55]{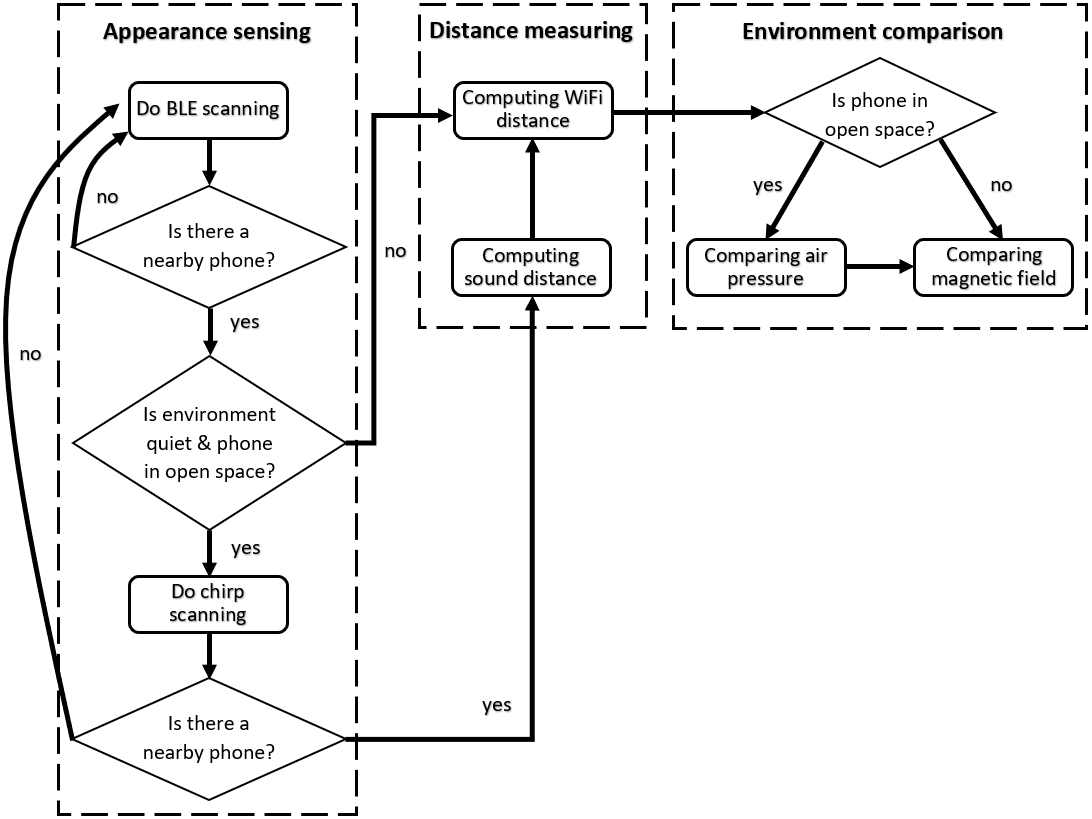}
    \caption{The flowchart of our contact tracing process which involves different smartphone sensors at each stage. At the end of each of the three main stages, an output regarding the phone's visibility, the relative distance estimate, and the environment difference are produced respectively.}
    \label{decision}
\end{figure}

Having studied the individual role of each sensor, we may now present our strategy to combine the above six sensors together. The system prioritises the appearance sensing of other nearby phones first, then moving on to measure the relative distance between them, and finally comparing the shared environment to reduce the number of false positives (see Figure~\ref{decision} and Table~\ref{sensorsemployed}). The detailed steps are as follows.

\begin{table}[h]
	\caption{Overview of the role of the smartphones sensors employed by our system.}
	\centering
	\begin{tabular}{c c c c}
		\toprule
		\textbf{Sensor} & \textbf{Appearance sensing} & \textbf{Distance measuring} & \textbf{Environment comparison} \\
		\midrule
		Barometer* & N/A & N/A & Shared air \\ \addlinespace[0.2cm]
		Bluetooth & Coarse-grained & N/A & N/A \\ \addlinespace[0.2cm]
		Magnetometer & N/A & N/A & Ambient magnetism \\ \addlinespace[0.2cm]
		Microphone & Fine-grained & Fine-grained & N/A \\ \addlinespace[0.2cm]
		WiFi & N/A & Medium-grained & N/A \\ \addlinespace[0.2cm]
		\bottomrule
	\end{tabular}
	
	* The proximity sensor is used as a trigger for barometer.
	\label{sensorsemployed}
\end{table}

\begin{description}[Type 1]
    \item[\textbf{Appearance sensing}]{\hfill \\
    \textit{Step 1:} The app periodically scans for nearby smartphones using the BLE signal. The rationale for using BLE as the primary sensor for appearance sensing, although other sensors (i.e. WiFi, microphone) are capable of the same function, is two-folds. Broadcasting distance wise, WiFi's range would be too long for consideration (with visibility of up to 50 metres away); acoustic sound, despite capable of adjusting its range, may not be reliable in noisy environment; whereas BLE offers the most compromised option for this matter. Power consumption wise, BLE consumes the least amount of energy amongst these sensors. This step is repeated until a smartphone is discovered.\\
    
    
    \textit{Step 2:} When a nearby phone is found by BLE, its range could be anywhere within 30 metres, including being on the other side of a thick wall or furniture. As such, this step attempts to further reduce these false positives with sound sensing. Firstly, the app does a preliminary check of the ambient noise. Since our chirps have an amplitude of around 20 $dB$, if the background noise is above this threshold, the system skips to Step 3, otherwise it performs a chirp scanning to see if the other phone can be heard (sound does not penetrate walls as easily as WiFi and BLE, hence there is a chance the phones are truly close).
    }
    
    \item[\textbf{Distance measuring}]{\hfill \\
    \textit{Step 3:} If the environment is too noisy, the system computes the WiFi based distance between the phones (although WiFi has a theoretical longer range than BLE, it does not use frequency hopping, hence resulting in much stable RSS). Otherwise, if the nearby phone can be reached by the chirp, the system computes the sound based distance, then proceeds to compute the WiFi based distance as well. In the end, those two distances are averaged for a better distance estimate.
    }
    
    \item[\textbf{Environment comparison}]{\hfill \\
    \textit{Step 4:} If both smartphones are in open space (as verified by the proximity sensor), the system compares their ambient air pressure, to assess whether the respective phone owners are breathing the same air (which is critical for airborne diseases). Otherwise, the system just compares the ambient magnetic field around the phones.
    }
\end{description}

The above processes may be viewed as in an on-line setting, where sensor measures arrive in real-time, and the system should decide on 3 output metrics, that are (in decreasing priority order), whether the two smartphones are close, what is the relative distance between them, and if they are sharing the same environment. Ideally, for a contact between the two phones to be registered, the app should detect that they are close (via BLE and/or sound), and their estimated distance is less than 1 metre according to WHO's guidance on infection (via WiFi and/or sound), and their shared environment is similar (via barometer and/or magnetometer). The final decision for each metric will be made over a 15 minute window (which is the duration to be infected according to the WHO's guidance), where the appearance can be a simple majority vote (e.g. if there are 10 samples in this period, in which over 50\% of them indicate a match, then the final decision is that they are close), the distance estimate and the environment sharing are an average of individual measures over this period window.

\subsection{Challenges}
Although the aforementioned sensors are undoubtedly useful for contact tracing, it is worth remembering that all of them were primarily designed for other purposes in mind. As such, the following challenges should be taken into account.
\begin{itemize}
    \item \textbf{Heterogeneous sensors.} Smartphones sensors come in different shapes and forms according to their makers, which may impact their sensitivities to the environment (e.g. some phones may receive a strong WiFi or BLE signal from far away thanks to a bigger antenna, whereas others need to be much closer). This challenge specifically impacts the environment observation approach, as one smartphone may view the world around differently from another phone. 
    
    \item \textbf{Noisy measures.} Smartphones sensors are miniaturised devices being packed tightly in a small phone body, whose measures are particularly noisy, not to mention the interference from other sources (e.g. the 2.4 GHz band is overcrowded with devices such as PC, laptop, microwave, radio, etc.). This challenge impacts the reliability of the sensors, as a sudden electronic noise may be misclassified as a true measure. 
    
    \item \textbf{Signal multi-path.} In a convoluted environment, there is rarely an unobstructed line-of-sight between the two smartphones. As such, the radio and sound waves travel in unexpected fashions in the air, which results in various receiving signals at the other end. This challenge strongly impacts the true distance estimate which is heavily based on the RSS. We address this concern by measuring several signals in the same place (i.e. increasing the probability of observing the good signal), and combining signals from different sensors. It is worth noting that as people approach closer to each other, there is less likely an obstacle between them, which lessens the impact of this challenge. 
\end{itemize}

\section{Empirical experiments}
\label{experiments}
Having presented our contact tracing proposal, we will now assess its feasibility and performance in various experiments. In doing so, we aim to address the following research questions.
\begin{itemize}
    \item Appearance sensing wise, what is the typical indoor and outdoor detection range of BLE and sound ?
    
    \item Relative distance wise, what is the offset between the estimated sound based, and WiFi based distance measure and the true distance ?
    
    \item Environment comparison wise, what is the feasibility of using air pressure and magnetic field ?
    
    \item Overall contact detection wise, what is the accuracy (in terms of the number of false positives) of our approach, compared to pure BLE based system ?
\end{itemize}

\subsection{Testbeds}
\label{sec:testbeds}
\begin{figure}[h]
	\centering
	
	\subfloat[The five-room office. There is plenty of electric appliances lying around (not shown here).]{\includegraphics[width=2.9in]{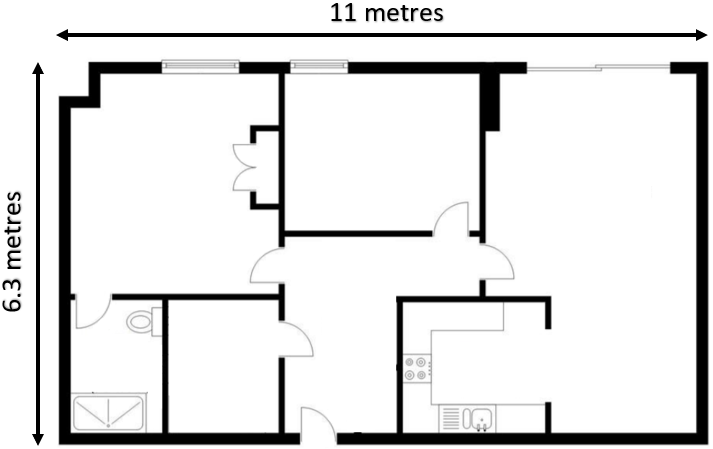}
		\label{officefloorplan}}
	\hspace{0.5pt}
	\subfloat[The ground floor of the 14-storey building. Only the communal areas (the shaded areas) were used.]{\includegraphics[width=2.6in]{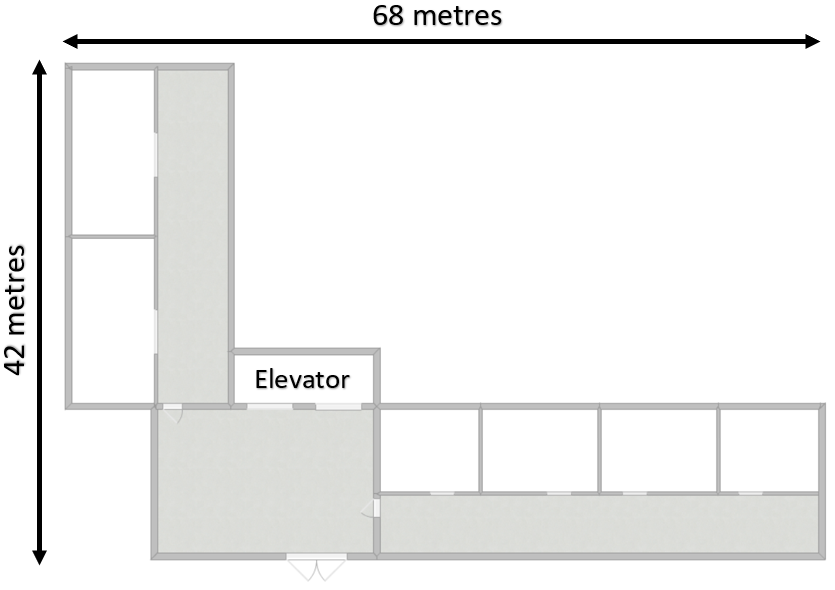}
		\label{groundfloorplan}}
	\hfil
	\subfloat[The parking garage. The shaded areas are the parking lots, which were mostly empty on the day. ]{\includegraphics[width=3.6in]{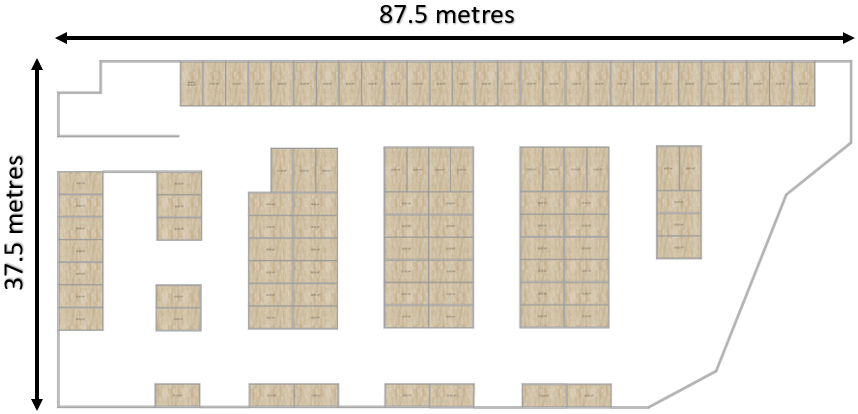}
		\label{parkingfloorplan}}
	
	\caption{The floor plans of the three testbeds.}
	\label{floorplans}
\end{figure}

Three realistic testbeds, representing both indoor and outdoor environment, were purposely selected to examine the feasibility and performance of our approach (see Figure~\ref{floorplans}). The first testbed contains a five-room office. The second testbed contains the communal areas of a 14-storey building. The third testbed is an open air parking garage.

Sample wise, for each testbed, we record 80 test instances, where each instance contains a pair of samples from two test phones at two different locations. In total, there are 240 test instances across the 3 testbeds. Distance wise, 60 of these instances are within 1 metre (i.e. the contagion range advised by WHO), another 60 of them are within 1 metre and 2 metres (i.e. the contagion range adopted by most European countries), 40 of them are within 2 metres and 3 metres (i.e. the false positive buffer zone), and the remaining 80 are somewhere between 3 metres and 30 metres (see Table~\ref{sampledistribution}).
\begin{table}[h]
	\caption{Overview of the distribution of test samples. More emphasis was given to samples within 3 metres as they are within the infection range of our interest.}
	\centering
	\begin{tabular}{c c c}
		\toprule
		\textbf{Distance between phones} & \textbf{Indoor samples} & \textbf{Outdoor samples} \\
		\midrule
		(0 - 1) metre & 40 & 20 \\ \addlinespace[0.2cm]
		(1 - 2) metres & 40 & 20 \\ \addlinespace[0.2cm]
		(2 - 3) metres & 20 & 20 \\ \addlinespace[0.2cm]
		(3 - 30) metres & 20 & 60 \\ \addlinespace[0.2cm]
		\bottomrule
	\end{tabular}
	\label{sampledistribution}
\end{table}

Ambient environment wise, the five-room office represents a typical indoor condition, with plenty of electric appliances (e.g. laptops, PCs, cameras) operating in the background, which may impact the wireless signal between the phones. The 14-storey building allowed our approach to be verified on different floor levels. The parking garage represents an ideal environment with wide open space containing unobstructed line-of-sight between the devices for minimal signal path loss. 


\subsection{Test devices and subjects}
\label{sec:devices}
Three smartphones were selected for testing, namely the LG G7 ThinQ, Samsung Galaxy S8, and Lenovo Phab 2 Pro (from now on, they will be simply referred to as LG, Samsung, and Lenovo phones). They were chosen to represent a variety of smartphones in the past 5 years, as well as covering different sensors manufacturers and Android operating systems (see Table~\ref{phones}). The LG phone will play the role as the central phone, whereas the Samsung and Lenovo phones will play the peripheral role.
\begin{table}[h]
	\caption{Overview of the 3 smartphones used to verify our approach.}
	\centering
	\begin{tabular}{c c c c c c c c}
		\toprule
		\textbf{Phone model} & \textbf{Year} & \textbf{Android} & \textbf{WiFi/BLE} & \textbf{Magnetometer} & \textbf{Microphone} & \textbf{Barometer} & \textbf{Proximity} \\
		\textbf{} & \textbf{released} & \textbf{OS} & \textbf{vendor} & \textbf{vendor} & \textbf{vendor} & \textbf{vendor} & \textbf{vendor} \\
		\midrule
		LG G7 ThinQ & 2018 & 9.0 & Qualcomm & Asahi Kasei & LG & LG & LG \\ \addlinespace[0.2cm]
		Samsung Galaxy S8 & 2017 & 7.0 & Murata & Asahi Kasei & Knowles & Samsung & Samsung \\ \addlinespace[0.2cm]
		Lenovo Phab 2 & 2016 & 6.0.1 & Qualcomm & Bosch & Lenovo & Lenovo & Liteon \\ \addlinespace[0.2cm]
		\bottomrule
	\end{tabular}
	\label{phones}
\end{table}

Using the above test devices, the sensor data were collected by two people at 240 test locations (as described in Section~\ref{sec:testbeds}), during daytime where there were other people around in the same premise, over the period of two months. During the experiments, the phones were either held in the user's hand or left in the pocket. For each test instance, we collected the sensor measures over 5 minutes (although a true positive contact is registered when two phones are within 1 metre for 15 minutes, according to the WHO's Covid-19 guidelines, yet, we relax the measuring time to speed up the experiments).

\subsection{Appearance sensing analysis}
\label{sec:appearance}
The purpose of the first experiment is assessing the visibility of BLE and sound, with varying distance between the two smartphones and the environment settings. For each of the 240 test locations, one smartphone constantly looks for the other during a fixed 1 minute period, keeping a record of each scan. We used the same pair of LG, Samsung phones in this experiment for consistency.

\begin{figure}[h]
    \centering
    \sidecaption
    \includegraphics[scale=.22]{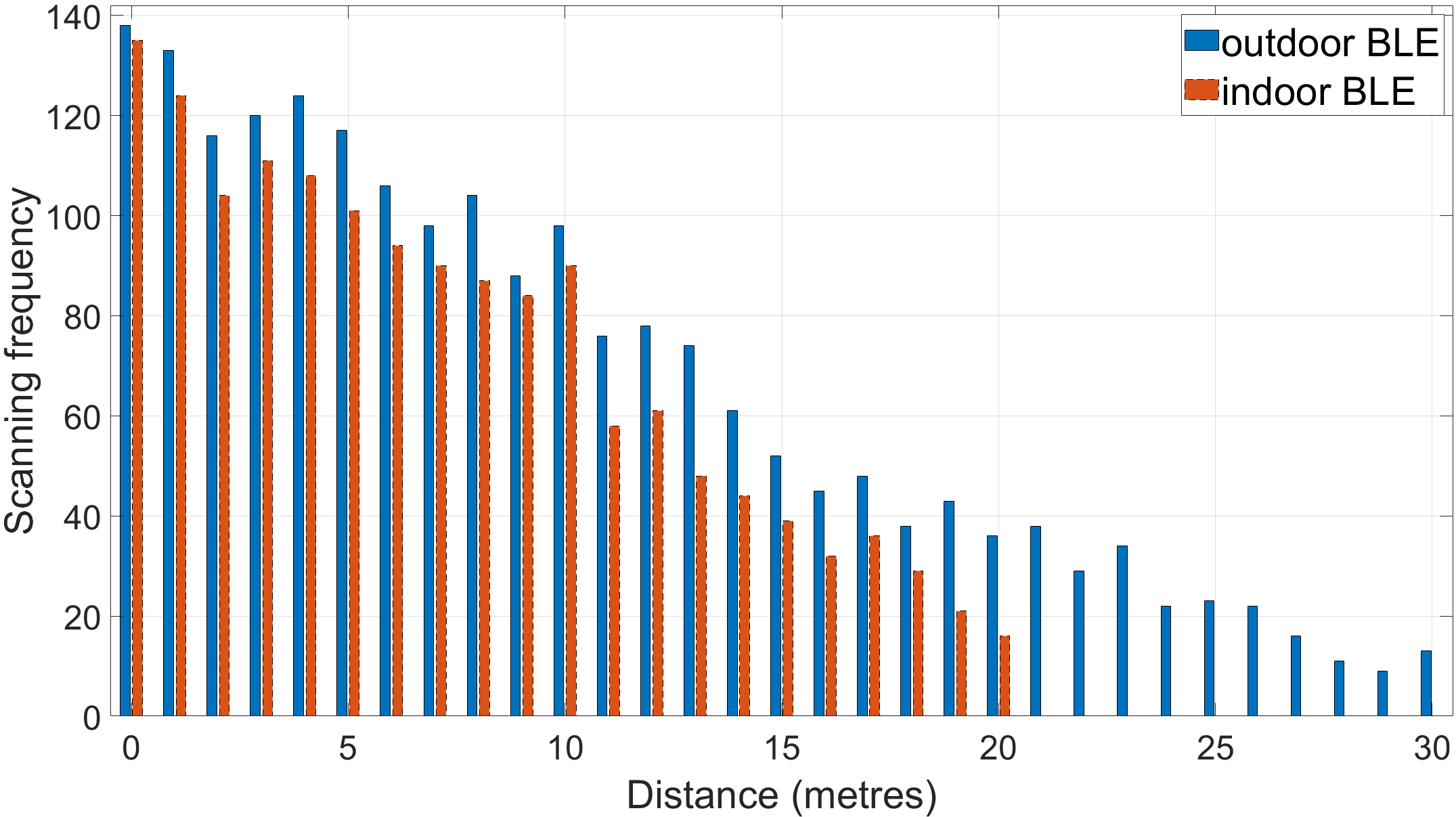}
    \caption{The scanning frequency of BLE, reported at 240 test locations. At each location, the central LG phone scans for 60 seconds and reports the result. In general, more measures were available outdoors than indoors, and at shorter distances.}
    \label{BLEvisibility}       
\end{figure}

For BLE, the result revealed three interesting trends (see Figure~\ref{BLEvisibility}). Firstly, the further the distance between of the two phones was, the lower the discoverability rate was. At 20 metres indoors, the detection number was just 16, compared to 124 of that when two phones were 1 metre apart. Secondly, the indoor visibility was noticeably poorer than outdoors (which is understandable with almost zero line-of-sight between the phones). Thirdly, they may still see each other at 20 metres apart indoors, and 30 metres outdoors. The third result has a strong implication for contact tracing, as these detections are clearly not in the contagion range, and may trigger a false positive.

\begin{figure}[h]
    \centering
    \sidecaption
    \includegraphics[scale=.17]{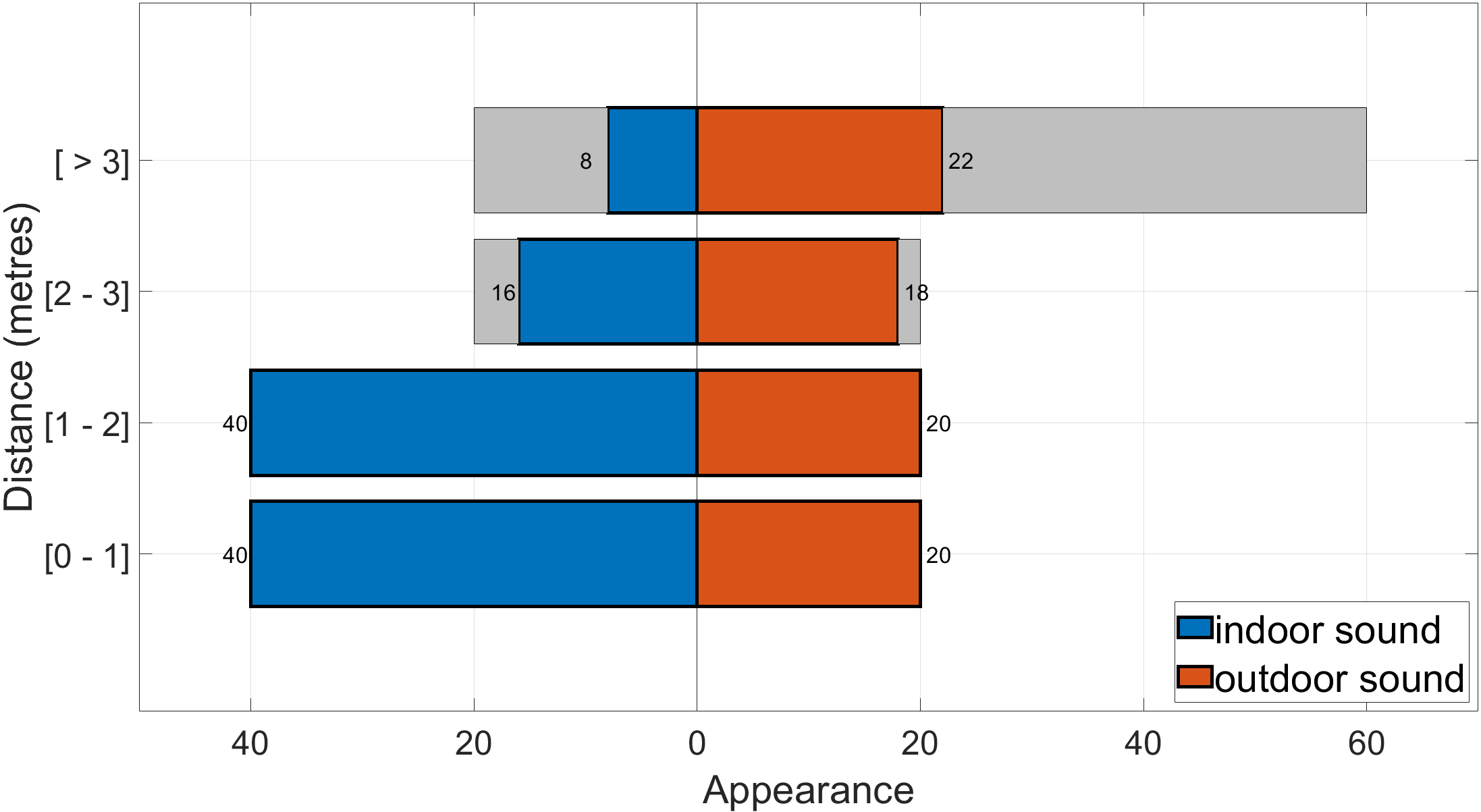}
    \caption{The hearing sensitivity of sound indoors and outdoors with respect to different distances between the two smartphones. The shaded areas represent test locations without any BLE or sound signal. Overall, the sensitivity decreases as the distance increases.}
    \label{indooroutdoorsound}       
\end{figure}

For sound, the appearance characteristics were similar to that of BLE, in which the further away the distance was, the less likely it could be heard. Yet, there was one clear distinction, that was, far fewer test instances beyond 2 metres can be heard with our chirp than with BLE (e.g. out of 20 test instances with more than 3 metres distance, only 8 of them could be reached with sound) (see Figure~\ref{indooroutdoorsound}).

\begin{figure}[h]
    \centering
    \sidecaption
    \includegraphics[scale=.17]{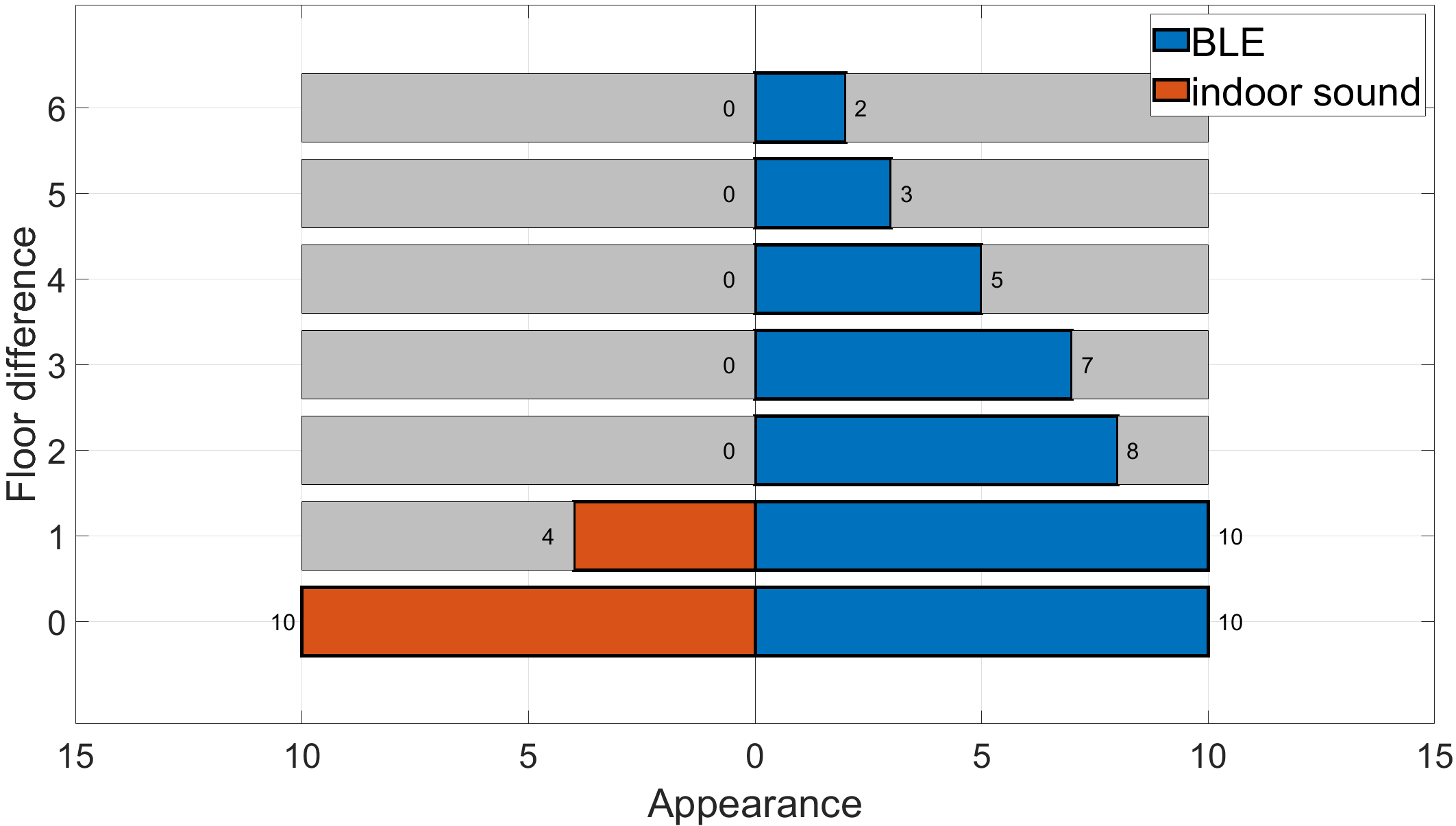}
    \caption{The visibility of BLE and sound on 13 floor levels, with 10 test locations per floor. The central LG phone was placed on the ground floor, while the Samsung phone keeps climbing up. No sound contact could be made beyond the 1$^{st}$ floor, where BLE can still be heard on the 6$^{th}$ floor. The shaded areas represent test locations without any BLE or sound signal.}
    \label{BLESoundFloor}       
\end{figure}

Since both BLE and sound may penetrate walls and furniture, their visibility was tested on different floor levels, with about 3 metre ceiling height. Interestingly, we got no sound feedback beyond the first floor, whereas BLE signal could penetrate up to the 6$^{th}$ floor (see Figure~\ref{BLESoundFloor}).

We do not report the scanning frequency for WiFi and sound, because of the WiFi scanning restriction on Android (at most 4 scans every 2 minutes), and the design of our chirp broadcasting interval (to avoid sound collision), as discussed in detail in Sections~\ref{WiFi} and~\ref{sound}.

In summary, the results in this section reveal that BLE-only based systems may score plenty more false positives due to its high visibility of up to 20 metres indoors and 30 metres outdoors. Sound, on the other hand, with much shorter travelling range, may offer an extra layer of appearance sensing information on top of BLE.

\subsection{Relative distance analysis}
\label{sec:distance}
This section assesses the feasibility and accuracy of the WiFi based and sound based distance estimate.

Since the only output we have, is just a simple number in the form of the WiFi RSS and sound pressure, we will assess how reliable this measure is, in the most ideal condition. As such, we performed the experiment in the open air parking garage, where the LG phone was fixed in one place, while the Lenovo phone moved away from it in a straight line. At every one metre along the way, 10 WiFi RSS and sound signal were taken. Theoretically, we would expect a steady decrease of such measures following the free-space path loss model, with respect to increasing distance.

\begin{figure}[h]
    \centering
    \sidecaption
    \includegraphics[scale=.17]{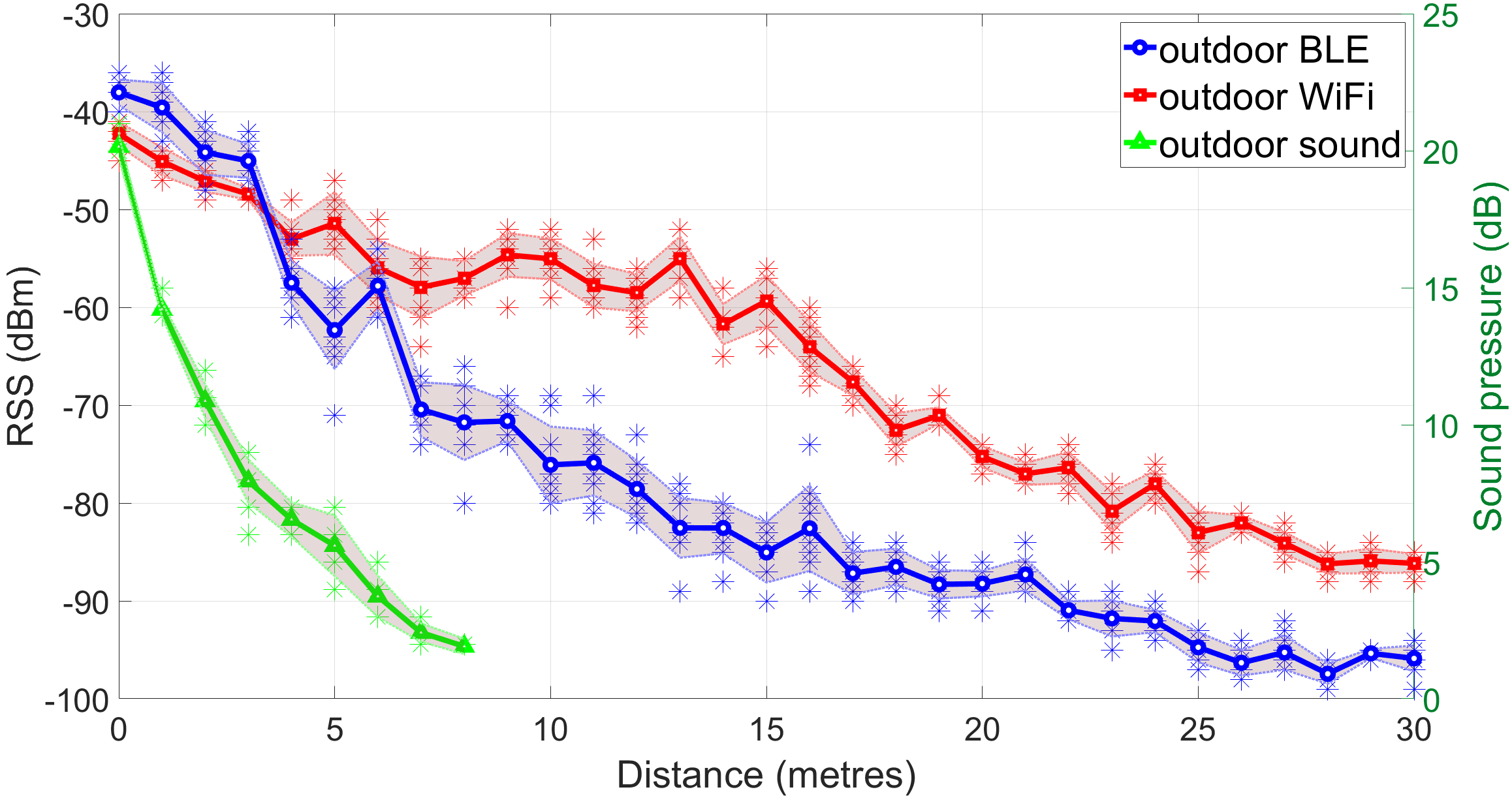}
    \caption{Comparison of the WiFi, BLE, and sound signal propagation in an ideal open air setting with a clear line-of-sight between the two smartphones. The sound signals are scaled by the right side y-axis. Acoustic sound and WiFi signals have the least amount of variations, compared to BLE, at the same location.}
    \label{WiFivsBLEvsSound}       
\end{figure}

For WiFi, the results revealed three surprising characteristics (see Figure~\ref{WiFivsBLEvsSound}). Firstly, even when both phones are static, with a clear line-of-sight between them and almost no interference from the environment (i.e. the signal may still bounce off the floor, nevertheless), the WiFi RSS still varies in the same spot. Secondly, strong signals (within 3 metres) and weak signals (beyond 20 metres) tends to be more stable than medium signals (from 3 metres to 20 metres). A plausible explanation is that at short distance the signal does not take long to travel, whereas by the time it travels 20 metres, it has lost most of its energy. Medium distance is the central spot for signal attenuation. Thirdly, at the same position, BLE RSS fluctuates significantly more than WiFi's, due to frequency hopping.

For sound, the result revealed a remarkably consistent measure with very little variation in this ideal outdoor setting. This ascertains its usefulness in estimating the relative distance between the smartphones.

Having understood the baseline expectation of BLE, WiFi and sound propagation characteristic, we may now assess their estimated distance's accuracy with our 240 test instances, both indoors and outdoors. For each test location, the true distance is the shortest straight line connecting the two smartphones discounting any walls and furniture in-between, the estimated WiFi and sound based distances are obtained from the free-space path loss model (as discussed in Sections~\ref{WiFi} and~\ref{sound}), the result is an average of 10 sound measures and 4 WiFi measures (the Android cap for 2 minute scan).

\begin{figure}[h]
    \centering
    \sidecaption
    \includegraphics[scale=.17]{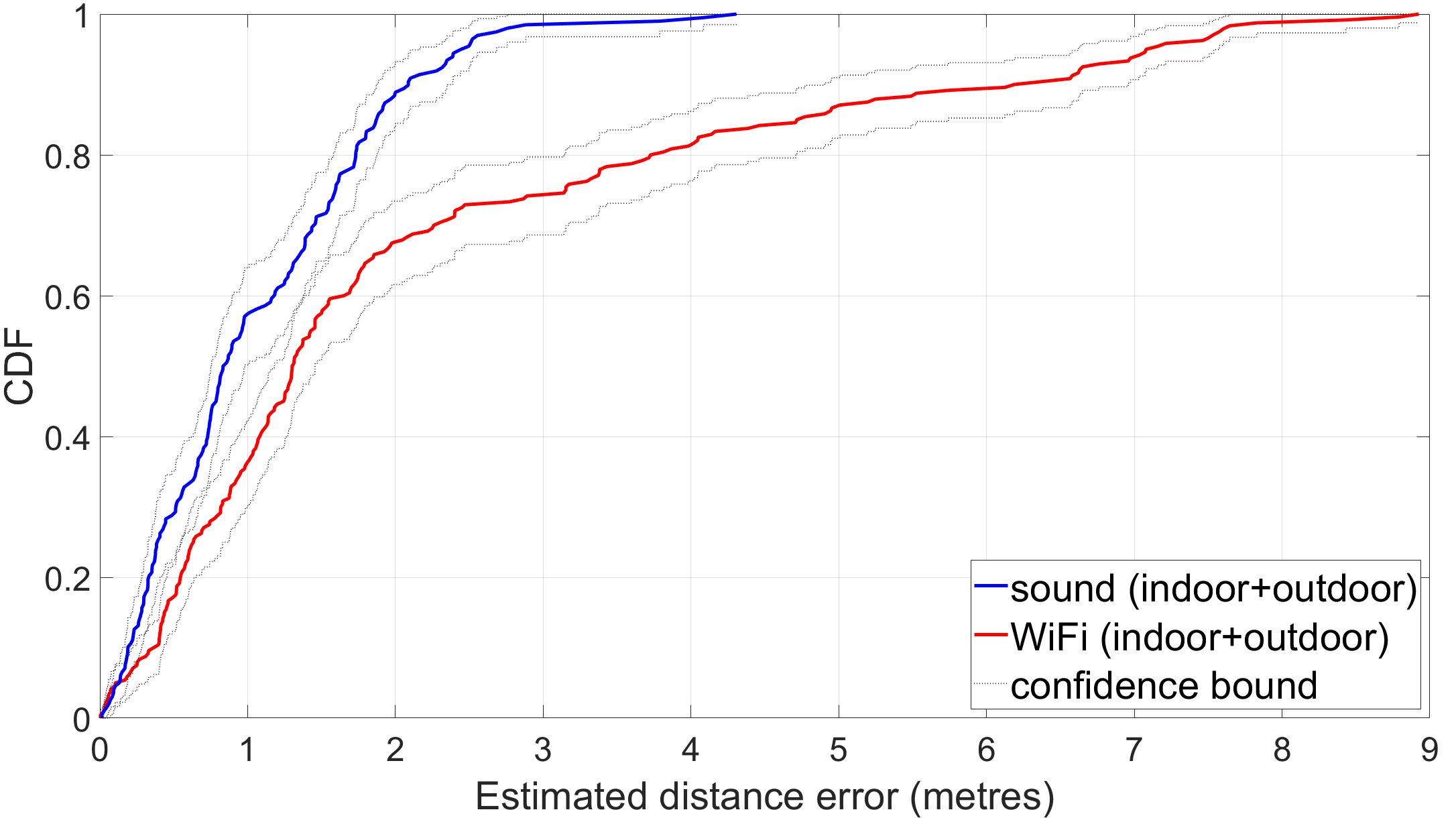}
    \caption{Comparison of the overall estimated distance based on WiFi and sound signal. The 95\% confidence bound of sound based result, calculated by the Kaplan-Meier estimator~\cite{goel2010understanding}, is noticeably tighter than WiFi one.}
    \label{CDFWiFiSoundOverall}       
\end{figure}

The result revealed the challenging impact of signal multi-path, where both WiFi based and sound based approaches produced high distance estimate error (see Figure~\ref{CDFWiFiSoundOverall}). In particular, sound based approach achieved less than 2 metre distance error 90\% of the time, whereas WiFi based approach could only manage 6.5 metre distance error 90\% chance, demonstrated by the Cumulative Distribution Function (CDF) plot. The confidence bound of sound based result, computed by the Kaplan-Meier estimator~\cite{goel2010understanding}, was also noticeably tighter than WiFi based one.

\begin{figure}[h]
    \centering
    \sidecaption
    \includegraphics[scale=.17]{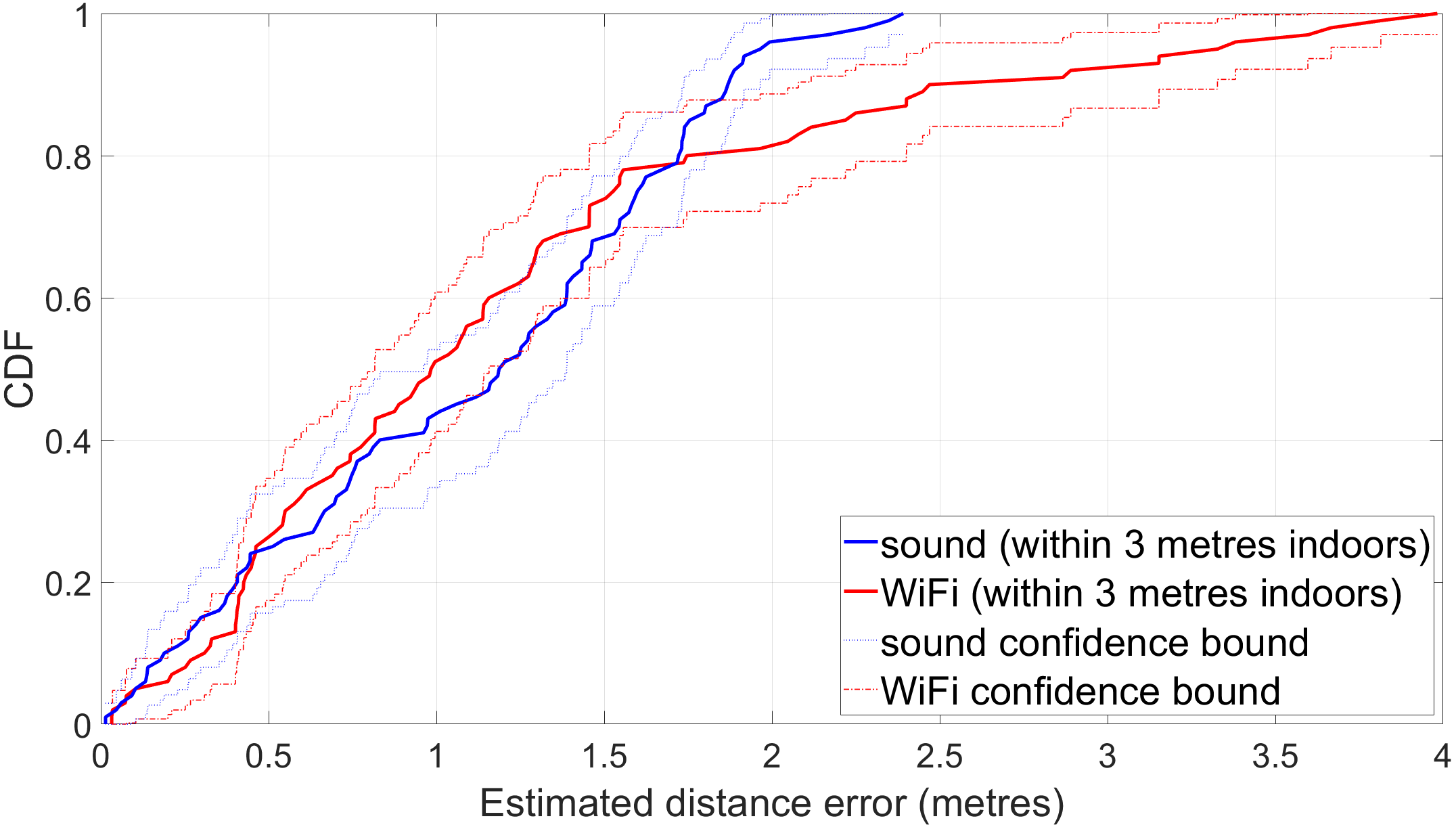}
    \caption{Comparison of the indoor estimated distance based on WiFi and sound signal for test locations less than 3 metres apart. WiFi based distance estimates were slightly more accurate than sound one.}
    \label{CDFWiFiSoundIndoor}       
\end{figure}

However, it was surprising that, when focusing on just indoor test locations within 3 metres, where there are furniture, walls between the smartphones, WiFi based estimated distances were slightly more accurate than sound one (see Figure~\ref{CDFWiFiSoundIndoor}). This result suggests that a combination of WiFi and sound may produce a better distance estimate.




\subsection{Environment comparison analysis}
\label{sec:environment}
This section assesses the feasibility and accuracy of using the barometer and magnetometer to determine whether the two smartphones are sharing the same environment, in particular, `breathing' the same air, which is critical for airborne infection.

Regarding air pressure, we assess its feasibility in close contact detection by experimenting its measure across the vertical and horizontal spaces.

\begin{figure}[h]
	\centering
	
	\subfloat[The air pressure on different floor levels exhibits a clear gap of around 0.43 $hPa$. The variation per floor is low at around 0.13 $hPa$.]{\includegraphics[width=3.3in]{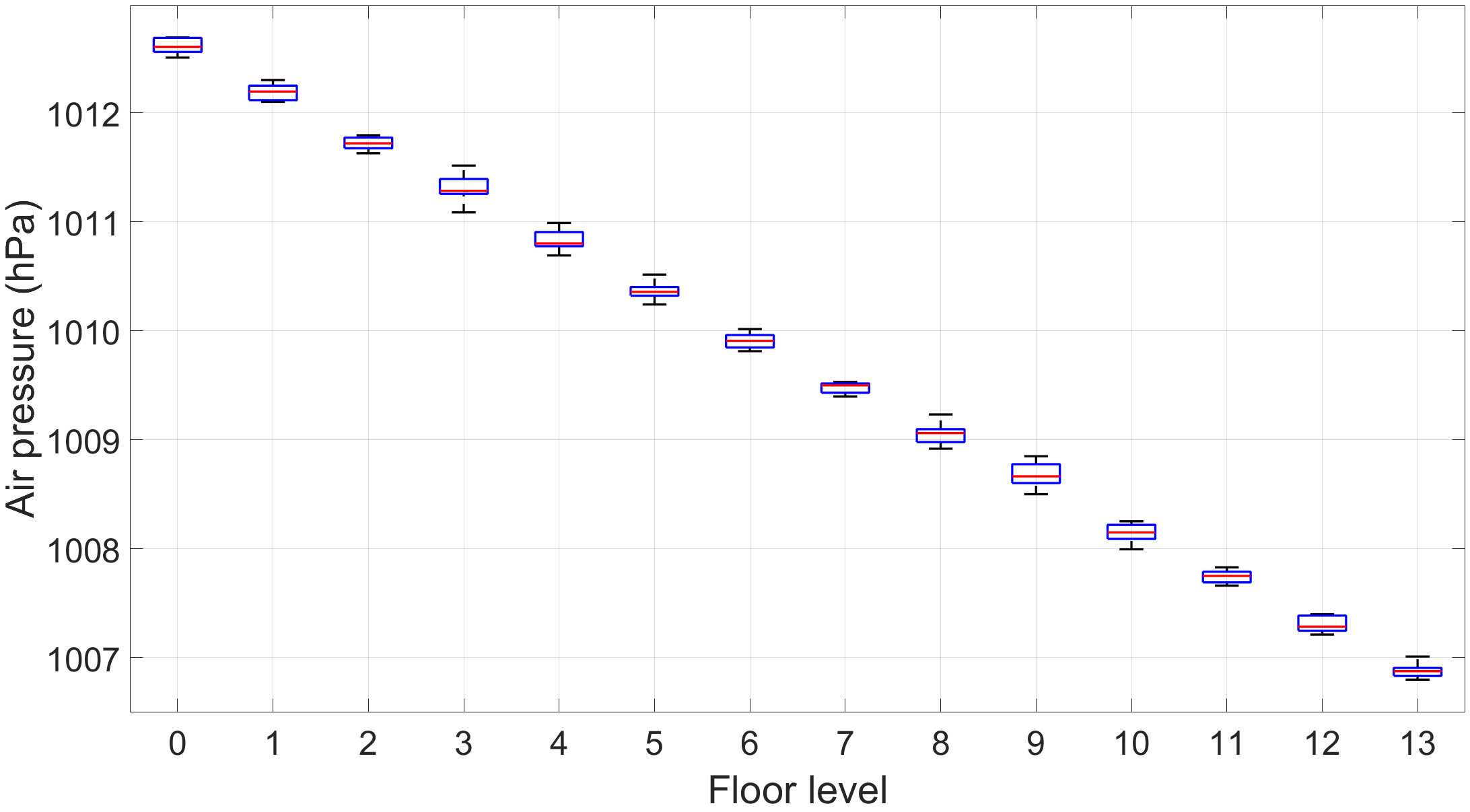}
		\label{pressurefloor}}
	\hspace{0.5pt}
	\subfloat[The distribution of the air pressure indoors and outdoors.]{\includegraphics[width=2.3in]{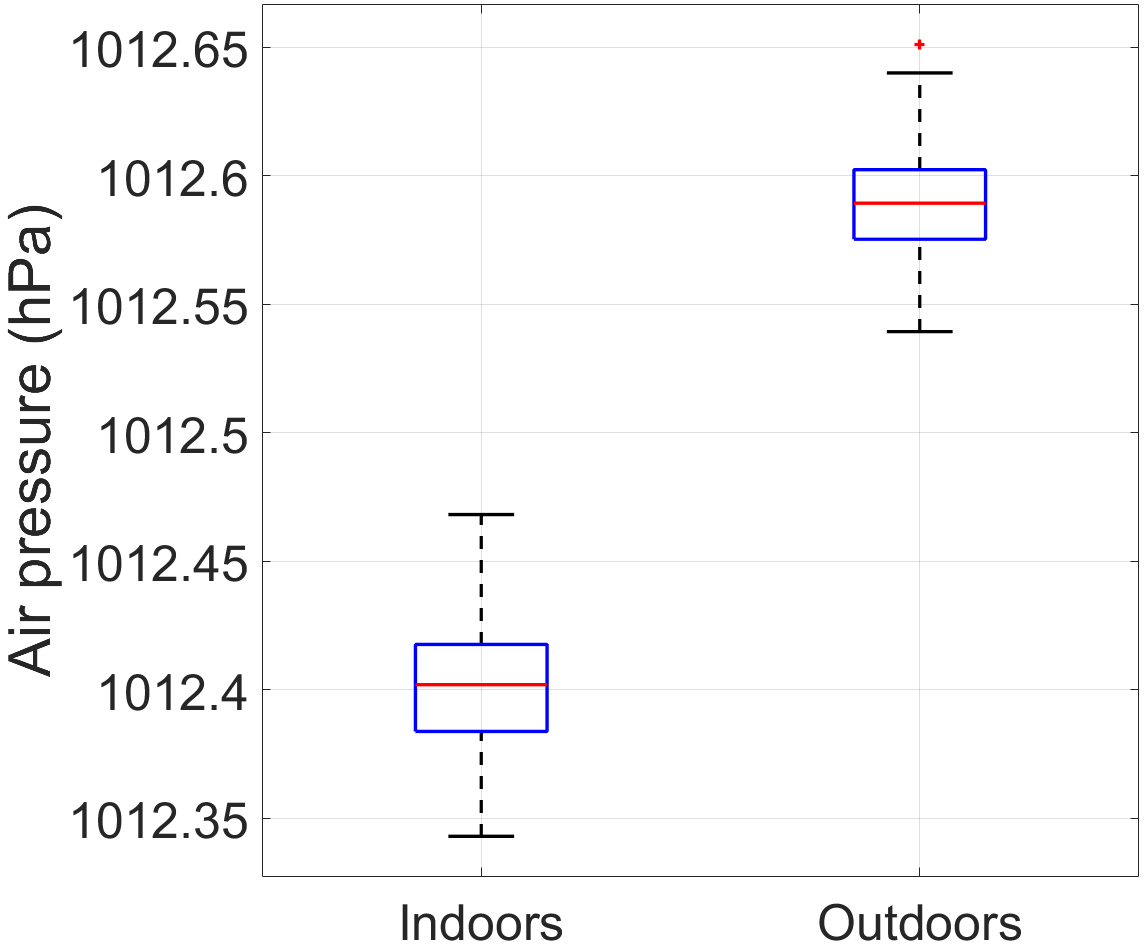}
		\label{pressureindooroutdoor}}
	
	\caption{The air pressure measures in different environments.}
	\label{airpressure}
\end{figure}

Vertical space wise, we collect 10 barometer readings with the LG and Samsung phones in the 14 floor building, with about 3 metre ceiling height separating each floor. The result reveals three useful information (see Figure~\ref{pressurefloor}). Firstly, the average air pressure strictly decreases as the altitude increases. Secondly, there was a clear measure gap of around 0.43 $hPa$ between each floor. Thirdly, the reading variations are particularly low (i.e. rather stable) per floor, of around 0.13 $hPa$. 

Horizontal space wise, we averaged the air pressure measures at 120 indoor test locations and 120 outdoor test locations, which reported 1,012.4 $hPa$ and 1,012.59 $hPa$ respectively (see Figure~\ref{pressureindooroutdoor}). The 0.19 $hPa$ average difference may provide a good indication to separate the indoor and outdoor users. Note that both the indoor and outdoor test environments are on pretty much the same ground level, with the parking garage locating next to the office building. The physical separation between them are walls, doors which may not be kept closed at all times. Therefore, a plausible explanation of such difference measure is that the building's exhaust ventilation system must have made an impact on the air pressure. Much more importantly, the variation of air pressures within the same building floor, or within the parking garage, is negligible and being lost within the sensor's mechanical noise, which turns out to be a great benefactor for our purpose, since smartphones within those areas are `breathing' the same air space.

\begin{figure}[!ht]
	\centering
	
	\subfloat[The five-room office exhibits high magnetic anomalies at various small areas.]{\includegraphics[width=2.9in]{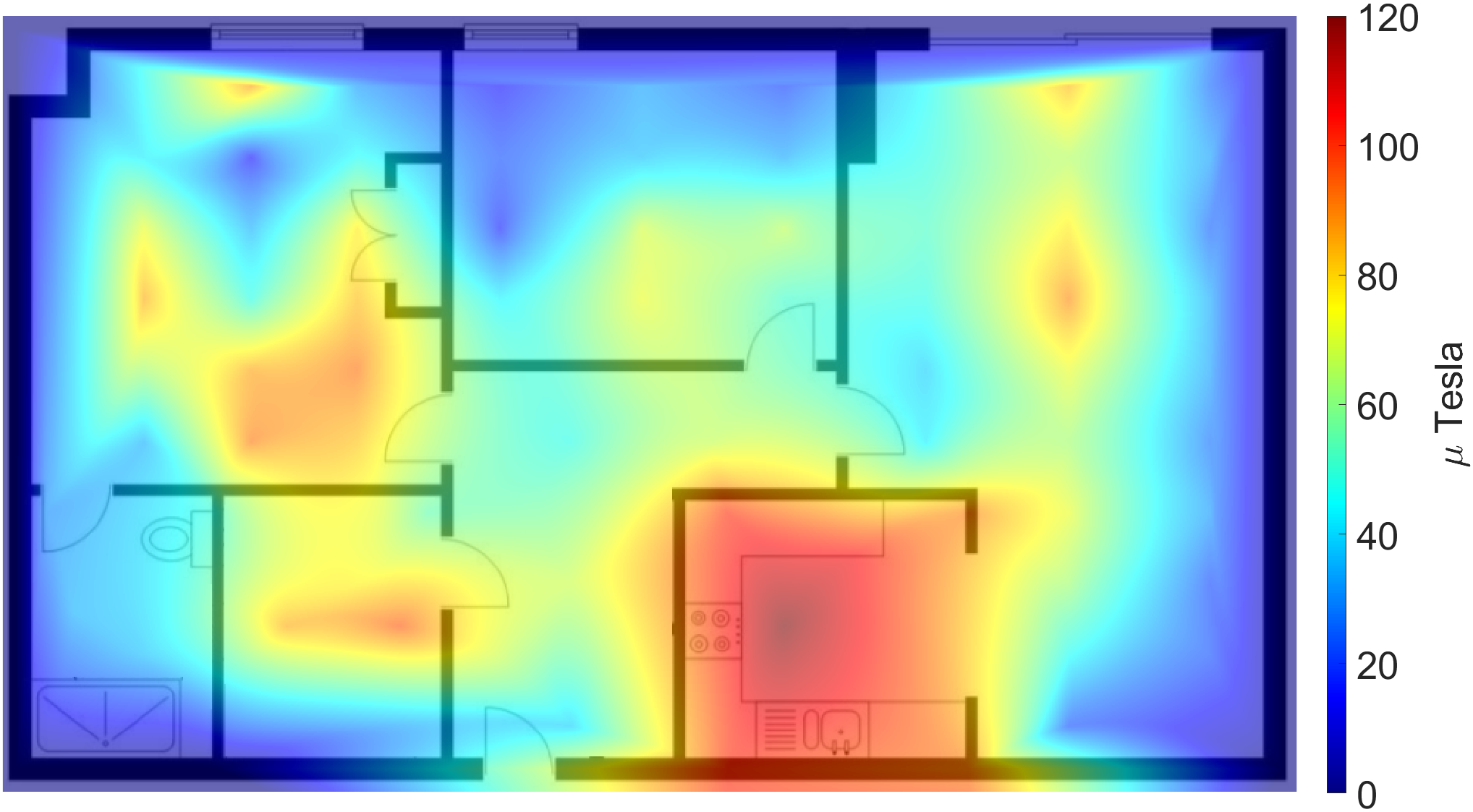}
		\label{heatmapoffice}}
	\hspace{0.5pt}
	\subfloat[Some communal spaces (e.g. by the elevator) on the ground floor of the 14-storey building contains some magnetic disturbances.]{\includegraphics[width=2.6in]{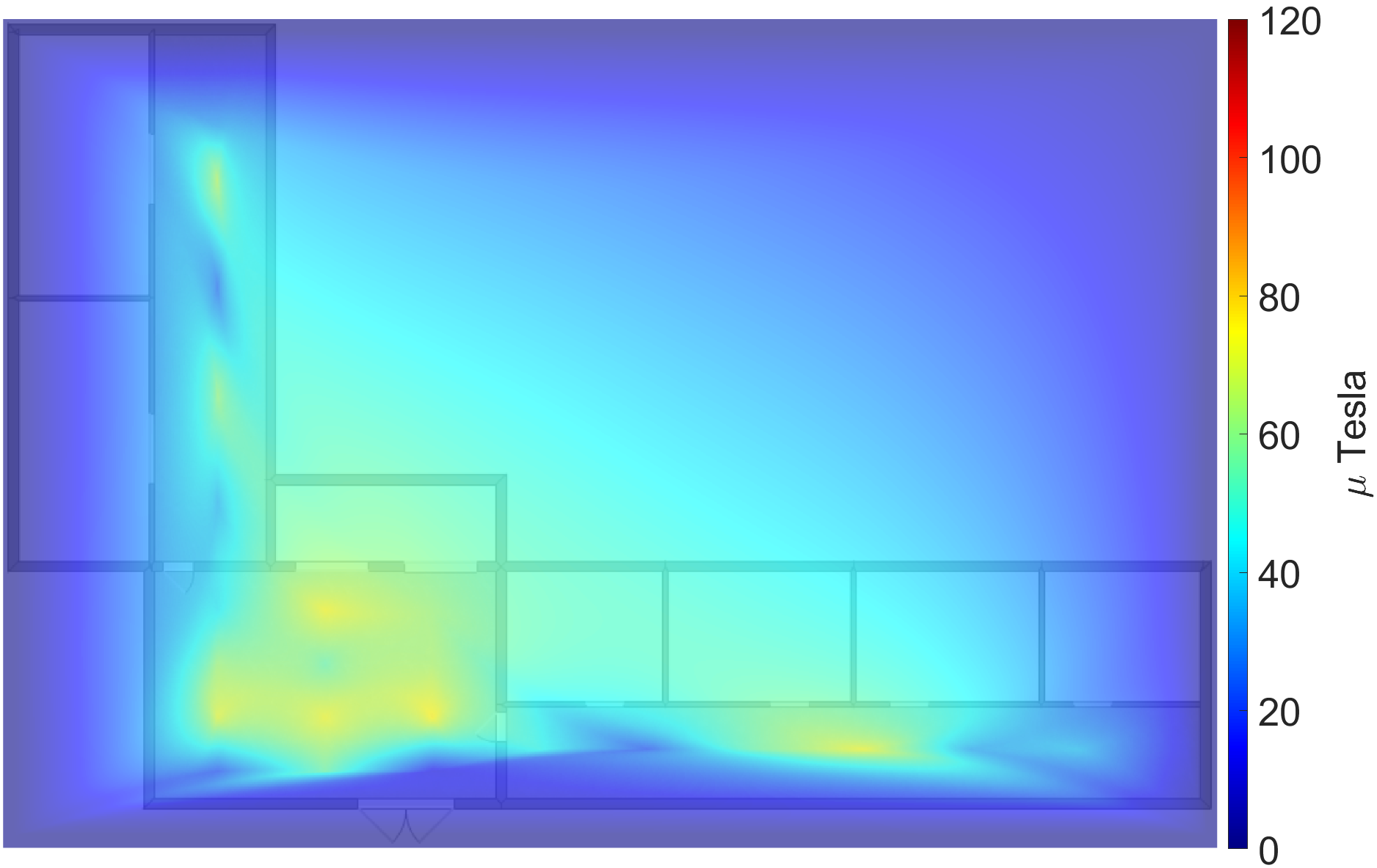}
		\label{heatmapgroundfloor}}
	\hfil
	\subfloat[The open air parking garage have low magnetic differences. There were not many cars around on the testing day.]{\includegraphics[width=4.1in]{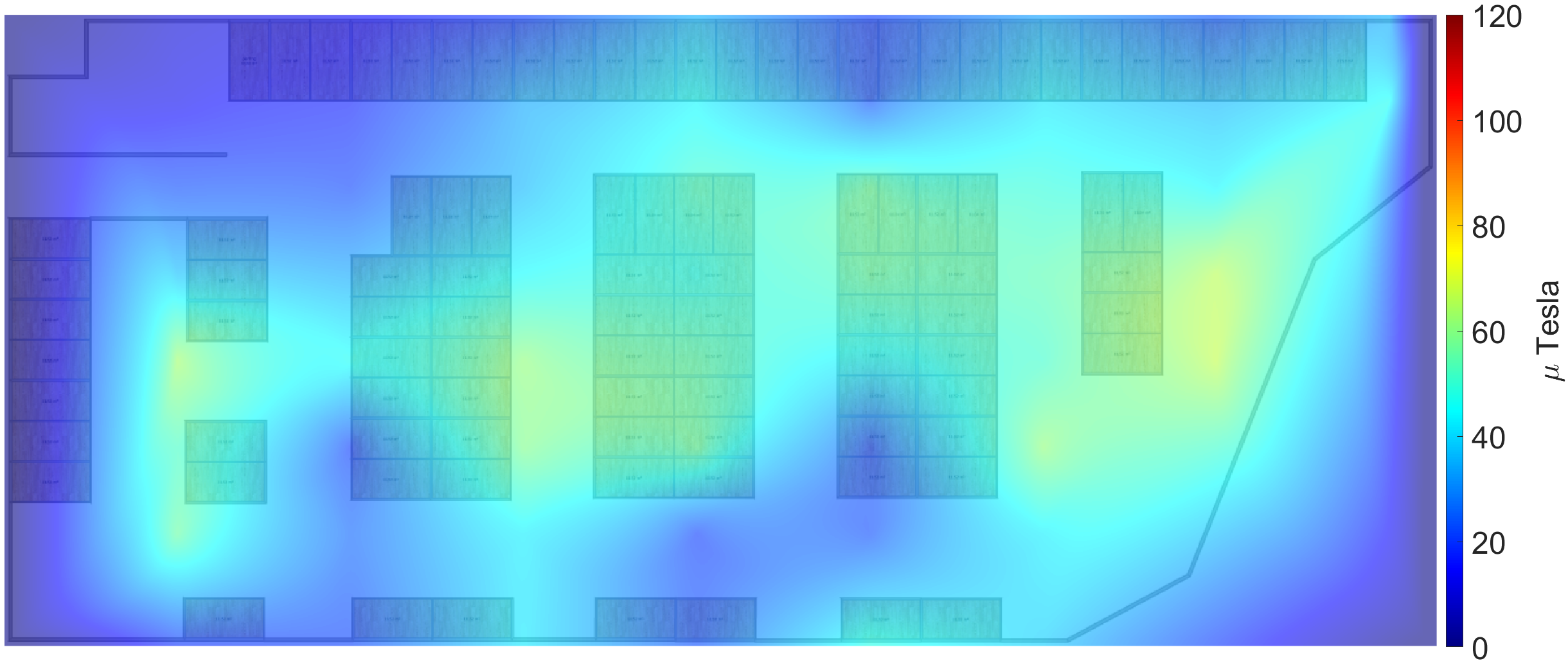}
		\label{heatmapgarage}}
	
	\caption{The magnetic heat map visualisation of the three testbeds. The colour maps have been put to the same scale.}
	\label{heatmaps}
\end{figure}

Regarding the magnetic field, the magnitude plot of 240 test locations across the 3 testbeds revealed two interesting observations (see Figure~\ref{heatmaps}). Firstly, the magnetic variation oscillates much stronger indoors than outdoors, with a maximum magnitude of nearly 120 $\mu T$ in the office testbed, compared to just 67 $\mu T$ in the parking garage testbed. The high number of electric appliances and the building materials may have been accounted for this variation. Secondly, the magnetic anomalies happened in small areas (i.e. only locations within a small room will have similar high magnetic disturbance).

\begin{figure}[h]
    \centering
    \sidecaption
    \includegraphics[scale=.17]{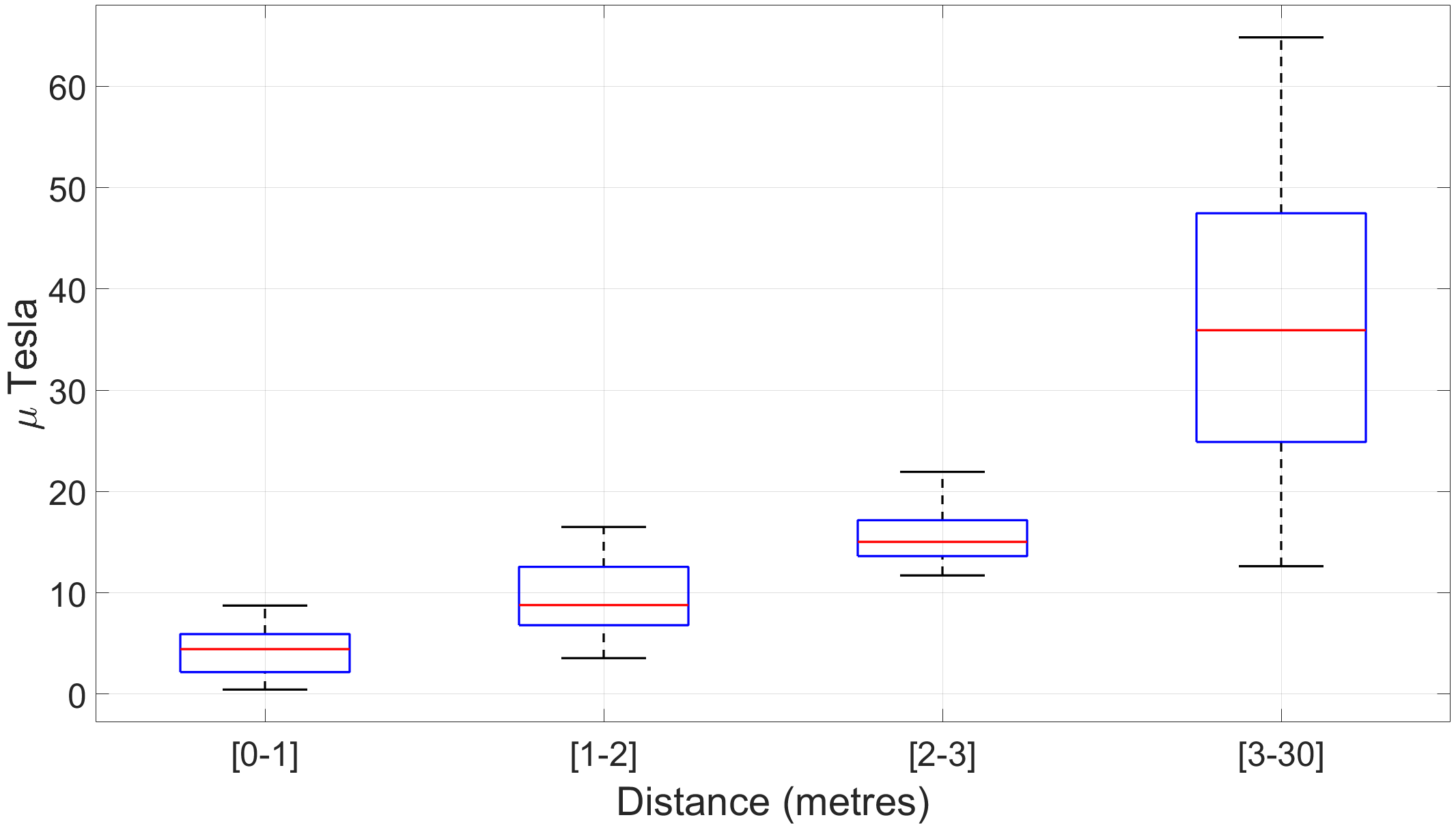}
    \caption{The Euclidean distance between the magnetic measures of the two smartphones at various distances. Close range samples have much smaller magnetic difference than long range ones.}
    \label{magneticdistance}       
\end{figure}

To assess the efficiency of the magnetic readings with respect to the distance between the two smartphones, we separate the 240 test locations into 4 categories, that are, tests within [0-1] metres, [1-2] metres, [2-3] metres, and [3-30] metres (more emphasis was given for those under 3 metres as they are of our particular interest), then computing the magnetic magnitude difference between the two smartphones in each category. The result demonstrated a much smaller Euclidean distance for close ranges, which implies that smartphones in increasingly long distance may observe much different magnetic readings (see Figure~\ref{magneticdistance}).

Ultimately, the question is, how would the air pressure and the magnetic field be employed to determine if the two smartphones are in close proximity ? Based on the above empirical experiments, air pressure wise, an empirical constant of 0.15 $hPa$ difference threshold (which is above the sensor noise, but is still below the air pressure separation) may be used to determine if two smartphones are in close proximity. Magnetic field wise, an empirical constant of 20 $\mu T$ difference threshold was chosen. In the next section, we will assess how those measures affect the overall system performance.


\subsection{Contact detection accuracy analysis}
\label{sec:accuracy}
Having assessed each of the three processes separately, we may now present the general system accuracy in detecting smartphones contact. The performance metrics are summarised in Table~\ref{metrics}.

For a true positive contact to be registered, both smartphones need to be within 1 metre of each other for at least 15 minutes (according to the WHO's Covid-19 infection guideline). In our tests, we will slightly neglect the time duration constraint, and will only observe 5 minutes per test location to speed up the experiment procedure. In contrast, a contact is considered as a false positive, if the phones were more than 1 metre apart (i.e. the contact should not have been registered by the system). Along the same line, a true negative contact is one which was not recorded by the system, and the phones were indeed more than 1 metre apart. Lastly, a false negative contact is one which was not recorded by the system, although the phones were within 1 metre in reality.
\begin{table}[h]
	\caption{Performance metrics in the contact tracing context.}
	\centering
	\begin{tabular}{c c}
		\toprule
		\textbf{Metrics} & \textbf{Note} \\
		\midrule
		True positive (TP) & True infection contact registered by the system \\ \addlinespace[0.2cm]
		False positive (FP) & Harmless non-infection contact registered by the system \\ \addlinespace[0.2cm]
		True negative (TN) & Harmless non-infection contact ignored by the system \\ \addlinespace[0.2cm]
		False negative (FN) & Actual contact but ignored by the system \\ \addlinespace[0.2cm]
		Overall accuracy & $\frac{TP + TN}{TP + TN + FP + FN}$ \\ \addlinespace[0.2cm]
		\bottomrule
	\end{tabular}
	\label{metrics}
\end{table}

Overall, out of 240 test instances, when only BLE appearance sensing was employed, the number of false positives was rather high at 180 (with most of them indoors) which results in a system accuracy of just 25\%. These figures are unsurprising given the high visibility of BLE. By incorporating distance measuring, the number of false positives dropped significantly to just 61, and more than doubled the system accuracy. However, the trade-off at this point was the increasing number of false negatives, which was unfortunately wrongly discarded by inaccurate distance estimates. Lastly, by applying the environment comparison on top, the number of false positives was further reduced to just 9, boosting the overall system accuracy to 87.08\% (see Table~\ref{overallperformance}).



\begin{table}[h]
	\caption{Performance comparison of different system mechanics. A true positive contact means the real distance between the smartphones is within 1 metre.}
	\centering
	\begin{tabular}{c c c c c c c}
		\toprule
		\textbf{System mechanics} & \textbf{Sensors employed} & \textbf{FP} & \textbf{FN} & \textbf{TP} & \textbf{TN} & \textbf{Accuracy} \\
		\midrule
		Appearance & (BLE only) & 180 & 0 & 60 & 0 & 25\% \\ \addlinespace[0.2cm]
		Appearance, distance & (BLE, WiFi, Microphone) & 61 & 22 & 38 & 119 & 65.42\% \\ \addlinespace[0.2cm]
		Appearance, distance, environment & (BLE, WiFi, Microphone, Barometer, Magnetometer) & 9 & 22 & 38 & 171 & 87.08\% \\ \addlinespace[0.2cm]
		\bottomrule
	\end{tabular}
	\label{overallperformance}
\end{table}

\subsection{Summary of results}
Having presented the above experimental results, we may now reflect on the research questions posed earlier.

With regard to sensing based on the device visibility only, two smartphones employing just BLE may still see each other well beyond the 2 metre contagious threshold (as employed by most European countries), in both indoor and outdoor environments with and without obstacles in-between. In some test instances, the BLE visibility could be up to 20 metres indoors and 30 metres outdoors. On the other hand, two smartphones more than 2 metres apart were barely communicable using sound (with a chirp amplitude of 20 dB). In particular, out of 20 indoor test instances beyond 3 metres, while all of them are visible with BLE, only 8 were reachable with sound. In addition, the further the distance between the phones was, the lower the discoverability rate was. This result suggests that BLE-only based system may return more false positives due to its high visibility, and several sensor measures should be taken at each location to ensure good coverage (see Section~\ref{sec:appearance} for more detailed results).

With regard to the relative distance estimate, it was highlighted that all technologies (i.e. WiFi, BLE, sound) struggled to convert their signal measures to precise distance estimation, especially indoors with a large gap between the devices, due to the signal multi-path phenomenon. However, when focusing on test locations within 3 metres, both WiFi and sound based distance estimates were around 2 metre error, 90\% of the time. This result suggests that WiFi and sound could be combined (e.g. averaged result) to produce a more accurate relative distance estimate (see Section~\ref{sec:distance} for more detailed results).

With regard to the feasibility of environment comparison, the experiments across 14 building floors revealed that the air pressure produced by the smartphone barometer strictly decreases as the altitude increases, with a clear measuring gap between each floor, and a low signal variation on the same floor. In addition, there was a clear difference between the indoor and outdoor air pressures. Lastly, the magnetic field measures vary significantly amongst small indoor areas. This result suggests that the air pressure and the magnetic field could provide an extra piece of information in helping to decide whether two smartphones are sharing the same airspace (see Section~\ref{sec:environment} for more detailed results).

With regard to the overall contact detection accuracy, for BLE-only system, the number of false positives was rather high at 180 (out of 240 test instances), with just 25\% accuracy. However, by incorporating WiFi and sound distance estimate, the system scored just 61 false positives, with more than 65\% accuracy. Finally, by applying air pressure and magnetic field comparison on top, the system only allowed just 9 false positives, with 87\% accuracy. This result demonstrated the usefulness of combining various smartphone sensors, over single BLE technology that most existing contact tracing works are currently relying on (see Section~\ref{sec:accuracy} for more detailed results).

\section{Conclusion and further work}
We have presented our novel, yet practical smartphone-based contact tracing approach. Through-out the six sections, we have thoroughly assessed the feasibility and the accuracy of our system in three realistic indoor and outdoor testbeds.

From our empirical results, we come to the conclusion that BLE only approach would suffer from a high number of false positives, due to its high visibility of up to 30 metres outdoors and 20 metres indoors, as well as high RSS variation because of its frequency hopping technique. WiFi signal, despite its long range distance, possesses much more stable RSS, and combining with sound signal would further enhance the distance estimate. More importantly, we recognise the natural property of WiFi, and BLE signals which may penetrate walls easily. To this end, air pressure measure provided by the barometer and the magnetic field magnitude provided by the magnetometer were employed to determine if the users are breathing the same air.

Ultimately, a system making vital decisions about the people's health such as contact tracing should be reliable and informative. Because of several number of parameters involved in the process (e.g. visibility, distance, time), it is more helpful to attach other information into all contact registrations (i.e. a contact estimated to be within 50 cm in more than 15 minutes should have a higher infection chance, than one estimated to be within 2 metres), rather than wholly relying on a binary (yes/no) contact detection. Additionally, exchanged information between the phones could be compressed to reduce the transferring time. This aspect is particularly relevant in the topic of information exchange and could further improve this work. Looking further ahead, it is critical to recognise that viral infections (e.g. Covid-19, SARS) are spread through the air. As such, contact tracing technologies should be able to make distinction between a true contact in shared air space, or a false one segregated by thick walls, to which, this paper is hoping to inspire future similar contact tracing approaches.



\begin{acknowledgement}
The authors are grateful to the three reviewers for their insightful comments that greatly improve this work. This research is partially supported by AstraZeneca.
\end{acknowledgement}

%
%

\bibliographystyle{spmpsci}
\bibliography{references}

\end{document}